# *Machine Learning Treasury Yields*

Zura Kakushadze[§†1] and Willie Yu[¶2]

[§] *Quantigic® Solutions LLC,[3] 680 E Main St, #543, Stamford, CT 06901*
[†] *Free University of Tbilisi, Business School & School of Physics*
*240, David Agmashenebeli Alley, Tbilisi, 0159, Georgia*
[¶] *Centre for Computational Biology, Duke-NUS Medical School*
*8 College Road, Singapore 169857*

January 6, 2020

Abstract

We give explicit algorithms and source code for extracting factors underlying Treasury yields using (unsupervised) machine learning (ML) techniques, such as nonnegative matrix factorization (NMF) and (statistically deterministic) clustering. NMF is a popular ML algorithm (used in computer vision, bioinformatics/computational biology, document classification, etc.), but is often misconstrued and misused. We discuss how to properly apply NMF to Treasury yields. We analyze the factors based on NMF and clustering and their interpretation. We discuss their implications for forecasting Treasury yields in the context of out-of-sample ML stability issues.

---

[1] Zura Kakushadze, Ph.D., is the President and a Co-Founder of Quantigic® Solutions LLC and a Full Professor in the Business School and the School of Physics at Free University of Tbilisi. Email: zura@quantigic.com

[2] Willie Yu, Ph.D., is a Research Fellow at Duke-NUS Medical School. Email: willie.yu@duke-nus.edu.sg

## 1. Introduction and Summary

It has been long appreciated that there is a structure underlying Treasury yields with different maturities. The common theme behind parametric factor models (e.g., [Nelson and Siegel, 1987], [Svensson, 1994], [Diebold and Li, 2006]), those based on principal components (e.g., [Litterman and Scheinkman, 1991], [Knez, Litterman and Scheinkman, 1994], [Bliss, 1997]) and others (see [Diebold and Li, 2006] for a literature review) is that the yield term structure is governed by a modest number (3 or 4) of underlying factors, which in some cases are identified as the level, slope (a.k.a. steepness) and curvature. Understanding the underlying factors is important both for bond portfolio hedging as well as possibly forecasting future Treasury yields.

In this paper we ask the following question: *Can we gain additional insight into the factors underlying Treasury yields by applying machine learning techniques, and can they aid with forecasting?* One immediate question that arises in this regard is which out of a plethora of machine learning techniques would make sense to apply to this problem. One glaring feature of the Treasury yields is that they are all nonnegative – at least for now. In this regard, nonnegative matrix factorization (NMF) [Paatero and Tapper, 1994], [Lee and Seung, 1999], which is an unsupervised machine learning algorithm (or, more precisely, a set of algorithms), would appear to be a natural candidate. Thus, if $Y$ is a matrix of yields (rows correspond to maturities and columns to dates), using NMF we can approximate it via $Y \approx W F$, where the nonnegative matrix $W$ (rows are labeled by maturities and columns by factors) is interpreted as the weights with which the $K$ factors encoded in the nonnegative matrix $F$ (rows are labeled by factors and columns by dates) contribute into the yields $Y$. It is the nonnegativity of $W$ that allows it to be interpreted as the weights (unlike in, e.g., the models based on principal components), whereas the nonnegativity of $F$ gives hope that there might be some underlying financial interpretation as the $K$ factors themselves are akin to yields, which can be appealing.

So, in Section 2 we apply NMF to Treasury yields and analyze the resultant factors.[4] NMF is a popular algorithm, but is often misconstrued/misused. NMF is a nondeterministic algorithm, so a single NMF run can produce a rosy-looking yet meaningless in-sample fit. In Section 2 we discuss a correct way of applying NMF (to Treasury yields) and provide R source code in Appendix A.[5] Our discussion there naturally leads us to an alternative approach based on statistically deterministic clustering, which we discuss in Section 3 and provide R source code in Appendix B. In-sample both the NMF and clustering approaches produce good fits and reasonably interpretable results. We then turn to the question of out-of-sample stability and forecasting, which we discuss in detail in Subsection 3.1. We briefly conclude in Section 4.

---

[4] NMF was applied to Brazilian yields data in [Takada and Stern, 2015], albeit without the nuances discussed below.
[5] The source code in Appendixes A, B and C hereof is not written to be "fancy" or optimized for speed or in any other way. Its sole purpose is to illustrate the algorithms we discuss. See Appendix D for some important legalese.

## 2. Nonnegative Matrix Factorization

We will organize Treasury yields into an $N \times T$ matrix $Y_{is}$, where $i = 1, \dots, N$ labels maturities, and $s = 1, \dots, T$ labels dates in the time series. Historical data for daily Treasury yield curve rates is freely available from https://www.treasury.gov/resource-center/data-chart-center/interest-rates/pages/TextView.aspx?data=yieldAll. The available maturities are 1 Mo, 2 Mo, 3 Mo, 6 Mo, 1 Yr, 2 Yr, 3 Yr, 5 Yr, 7 Yr, 10 Yr, 20 Yr and 30 Yr. The data is available starting January 2, 1990. There are N/As in the data. The 1 Mo maturity data is available from July 31, 2001. The 2 Mo maturity series begins on October 16, 2018. The 20 Yr maturity series was discontinued at the end of 1986 and reinstated on October 1, 1993. The 30 Yr maturity series was discontinued on February 18, 2002 and reintroduced once again on February 9, 2006.

Nonnegative matrix factorization (NMF) [Paatero and Tapper, 1994], [Lee and Seung, 1999] approximates the nonnegative $N \times T$ matrix $Y_{is}$ via $Y \approx W F$, where the columns of the nonnegative $N \times K$ matrix $W_{iA}$ ($A = 1, \dots, K$) are interpreted as the weights with which the $K$ factors (or "exposures") encoded in the nonnegative $K \times T$ matrix $F_{As}$ contribute into $Y_{is}$. The weights and factors can always be normalized such that the columns of $W_{iA}$ add up to 1:

$$\sum_{i=1}^{N} W_{iA} = 1 \qquad (1)$$

The number $K$ of factors is a hyperparameter, which is usually fixed by finding the best (in-sample) fit. However, below we will also use an independent methodology for inferring $K$.

A nice thing about using NMF in the context of yields is that, since the weights and factors are nonnegative, a priori we can hope to obtain a clear financial interpretation of the underlying factors. However, the flipside is that NMF is nondeterministic. That is, an NMF algorithm does not isolate the global minimum (when minimizing an appropriately defined error function).[6] Instead, in each run, which is seeded randomly, NMF finds one out of a large number of local minima. So, while each individual NMF run can produce a very rosy-looking in-sample fit (which is a common pitfall), even in-sample this means very little as different runs can produce substantially different-looking results. One way of dealing with this is to average over a sizable number $P$ of runs and look at not only the average value, but also the error.[7]

Here is one way to do this. Each run labeled by $r = 1, \dots, P$ produces the corresponding weights and factors matrices $W^{(r)}$ and $F^{(r)}$. The averaged weights matrix $W$ can be defined, element by element, as the mean (or median) value from the $P$ runs. The averaged factors matrix $F$ can be defined similarly. However, each element now has an error, which can be defined as the standard deviation (or MAD = mean absolute deviation) across the $P$ runs. If we define the fitted matrix $\tilde{Y} = W\,F$, where $W$ and $F$ are the averaged matrices defined as above, then the resultant fit of $\tilde{Y}$ to $Y$ may not be so rosy. Furthermore, the error bars for $W$ and $F$ can be substantial, thereby obscuring any financial interpretation gained by using NMF. So, care is needed when using NMF, and below we will discuss how to reduce the resultant noise.

---

[6] The quantity (to be minimized) that measures the fit in this approximation can be defined as the sum (over both $i$ and $s$) of the squares of the element-by-element errors (which is the Frobenius norm of the error matrix $Y - W F$).
[7] It appears that this important point was not addressed in [Takada and Stern, 2015].

Averaging over multiple NMF runs discussed above may appear straightforward, but there is a complication. Let us focus on the weights matrix as the issue with the factors matrix is similar. Each run labeled by $r = 1, \ldots, P$ produces the $N \times K$ matrix $W^{(r)}$. However, the columns of these $P$ matrices from different runs are not aligned. We must align them before averaging, which is nontrivial. We can use clustering to align them. This can be done as follows. Let us bootstrap the $P$ matrices $W^{(r)}$ column-wise into the $N \times (P \cdot K)$ matrix $\widehat{W}$. We can now cluster the $(P \times K)$ columns of $\widehat{W}$ (each column being an $N$-vector) into $K$ clusters using k-means [Lloyd, 1957], [Steinhaus, 1957], [Forgy, 1965], [MacQueen, 1967], [Hartigan, 1975], [Hartigan and Wong, 1979], [Lloyd, 1982]. This way we can map each column of each matrix $W^{(r)}$ to the set $1, \ldots, K$, thereby aligning them, so we can now average over them.

However, there are some possible "hiccups" with the alignment of the columns of the matrices $W^{(r)}$ described above. First, a priori there is no guarantee that the $(P \cdot K)$ columns of $\widehat{W}$ will be mapped into precisely $K$ batches with $P$ elements in each batch. Generally, we could have $K$ batches with $P_1, \ldots, P_K$ elements, where some $P_A$ ($A = 1, \ldots, K$) are different from $P$. Now, a priori this is not necessarily problematic when $P_A$ are all large. In this case we can simply average over the $P_A$ columns in each batch to arrive at the $A$-th column of $W$, albeit the fact that we get nonuniform $P_A$ can be indicative of an instability (see below). A more pressing issue arises when $P$ is small, to wit, some $P_A$ can be 0, which implies that our guess for $K$ overshoots and the actual number of factors is smaller. In fact, one can argue that if $P_A$ is 1 (or smaller than some predefined threshold), the same conclusion applies. This can dealt with simply by (repeatedly) reducing the number of factors from $K$ to $K - 1$ and reapplying the above procedure. Finally, k-means is itself a nondeterministic algorithm, so given the same set of $P$ matrices $W^{(r)}$, the clustering of their columns can be different from one k-means run to another. This is not necessarily problematic for two reasons. First, assuming that the input data $Y_{is}$ is not completely random and there is indeed some underlying factor structure in it, the degree of such nondeterminism is substantially lower than the nondeterminism inherent in NMF and can be expected to be (well) within the error bars resulting from averaging over the $P$ matrices $W^{(r)}$ once they are aligned. Second, the k-means nondeterminism, as we will see below, is related to the noise in the $P$ matrices $W^{(r)}$, and once we reduce this noise using a method we discuss below, the k-means nondeterminism is also greatly reduced (or disappears).

## 2.1. What Is the Number of Factors?

So, what should we take as the number of factors $K$? One way is to keep it as a hyperparameter and fix it by trial and error, i.e., by identifying the value of $K$ for which we get the best in-sample fit. For this we need to define a measure of what a good fit is. We will come back to this point below. However, here we will pursue a more holistic approach to fixing $K$.

The idea for fixing $K$ we discuss here is not new and was used in the context of statistical risk models in [Kakushadze and Yu, 2017a] and cancer signatures in [Kakushadze and Yu, 2016a]. Consider the $N \times N$ *serial* correlation matrix ($i, j = 1, \dots, N$)

$$\Psi_{ij} = \text{Cor}(Y_{is}, Y_{js}) \tag{2}$$

We can infer an effective dimensionality of this matrix via eRank (or effective rank) [Roy and Vetterli, 2007], which is defined as the exponential of the Shannon a.k.a. spectral entropy [Campbell, 1960], [Yang, Gibson and He, 2005].[8] Using the Treasury yield data (see above) for the period October 16, 2018 through November 22, 2019 (both inclusive),[9] we get eRank($\Psi_{ij}$) ≈ 1.43. So, based on this, we expect 1 or 2 relevant factors. However, once we dig deeper, we find that this result can be a bit misleading and can be improved. The issue here is that the average pairwise correlation (i.e., the average over the $N(N-1)$ values $\Psi_{ij}$ with $i \neq j$), unsurprisingly, is whopping 88.82%. This is because typically yields at different maturities are highly correlated, so we have an "overall mode"[10] corresponding to the average pairwise correlation and essentially governed by the first principal component of $\Psi_{ij}$. As a result, the eigenvalue corresponding to the first principal component is much larger than other eigenvalues of $\Psi_{ij}$ and contributes into eRank($\Psi_{ij}$) with a dominant weight. To circumvent this, we can drop the first principal component from $\Psi_{ij}$ thereby obtaining a new matrix $\Psi'_{ij}$ and define the modified eRank as ModeRank($\Psi_{ij}$) = eRank($\Psi'_{ij}$) + 1 [Kakushadze and Yu, 2017a]. The so-defined modified eRank is a better measure of the effective dimensionality of $\Psi_{ij}$ than the vanilla eRank. For our dataset we get ModeRank($\Psi_{ij}$) ≈ 2.34. So, we expect 2 or 3 relevant factors, which is consistent with other approaches and our actual results (see below).

An important lesson learnt from analyzing the serial correlation matrix $\Psi_{ij}$ is that we have the dominant "overall mode". In the context of cancer signatures [Kakushadze and Yu, 2016a] this makes vanilla NMF noisy. We will see below that the same transpires with yields.

---

[8] R source code for computing eRank (with and without the first principal component – see below) is given by the subfunction `calc.erank()` in the function `qrm.erank.pc()` in Appendix A of [Kakushadze and Yu, 2017a].
[9] We downloaded the data on November 24, 2019, hence the end-date of the data. The start date is when the 2 Mo maturity series began (see above), so this period conveniently contains all twelve maturities mentioned above.
[10] A similar "overall mode" is observed in the context of cancer signatures [Kakushadze and Yu, 2016a]. In the context of a broad equities basket, this "overall mode" is known as the "market mode" (see, e.g., [Bouchaud and Potters, 2011], [Kakushadze and Yu, 2017a]), which corresponds to the overall movement of the "broad market".

## 2.2. Vanilla NMF

Let us now apply vanilla NMF (without any de-noising, which we discuss below) to our dataset (spanning the period October 16, 2018 through November 22, 2019). We present the results averaged over $P = 100$ runs (see above), albeit practically speaking one can get similar results with fewer runs. However, since the dataset is not very large, the code (which we give in Appendix A) runs fast enough, so $P = 100$ is by no means computationally taxing. When combining the $P$ runs, we use mean and standard deviation (as opposed to median and MAD).[11]

As a measure of how good a fit we get, we use two numbers for each maturity. For each row of the original data matrix $Y$ we calculate its *serial* correlation with the corresponding row of the fitted matrix $\tilde{Y} = W\,F$. This way we obtain $N$ correlations $\rho_i = \text{Cor}(Y_{is}, \tilde{Y}_{is})$. For each maturity, we also compute the serial sum of squares:

$$E_i = \sum_{s=1}^{T} (Y_{is} - \tilde{Y}_{is})^2 \tag{3}$$

The results for $K = 2$ are given in Tables 1-2 and Figures 1-4. The results for $K = 3$ are given in Tables 3-4 and Figures 5-10. From these results it is evident that vanilla NMF is very noisy, with large error bars. Furthermore, adding the third factor does not improve the fit or reduce the noise. In fact, upon a closer examination, it is clear that the first factor for $K = 3$ corresponds to the first factor for $K = 2$, the third factor for $K = 3$ corresponds to the second factor for $K = 2$, while the second factor for $K = 3$, which is extremely noisy, is new compared with $K = 2$. This new factor has every characteristic of the level (in the "level", "steepness" and "curvature" nomenclature of the factors underlying yield curves – see below), except that it is so noisy that it can only be taken with a grain of salt. Nonetheless, the hint that we have the level present in the data, and that it is noisy, is very useful. In fact, we anticipated its presence in the previous subsection when we analyzed the correlation matrix $\Psi_{ij}$. The level is related to the dominant "overall mode" we discussed above. Generally (and not necessarily in the context of NMF), the level can be defined as the factor $F_{As}$ for which the corresponding weights $W_{iA}$ are uniform: $W_{iA} \equiv 1/N$ (in the normalization of Eqn. (1), which is convenient in the context of NMF; in other contexts, equivalently, one would often set $W_{iA} \equiv 1$). Hence the high correlations between different maturities. The noisiness of the level then is propagated to the other factors and enhances their noisiness.[12] This has already been discussed in the context of cancer signatures in [Kakushadze and Yu, 2016a], whose solution to noisiness we apply here.

---

[11] The latter give smaller error bars for the weights, but not necessarily for the factors, and do not produce a better overall fit. Also, the error bars are reduced dramatically with de-noising we discuss below, so this is a moot point.
[12] Conceptually, this is similar to the noisiness of the intercept factor (whose factor loadings are uniform) in a cross-sectional linear regression of stock returns, e.g., in the context of serial t-statistic [Fama and MacBeth, 1973].

*2.3. De-Noised NMF*

The idea of [Kakushadze and Yu, 2016a], which we adapt here with appropriate tweaks, is simple: factor out the noisy "overall mode" before applying NMF. This is what we refer to as de-noising the original data $Y_{is}$. However, in the context of cancer signatures [Kakushadze and Yu, 2016a], the corresponding data is comprised of mutation counts with roughly log-normal distributions and an exponential structure, so de-noising the data there essentially amounts to taking the log of the original data matrix, demeaning it cross-sectionally, and then re-exponentiating, which results in a positive matrix. The yields $Y_{is}$ do not possess such an exponential structure, so we cannot apply the procedure of [Kakushadze and Yu, 2016a] here directly. Also, simply demeaning $Y_{is}$ cross-sectionally will not work as the so-demeaned matrix will not be nonnegative thereby defying the purpose of NMF. So, we must find another way of de-noising the matrix $Y_{is}$ such that it would essentially amount to factoring out the level factor.

Happily, there is a simple solution to this. For each value of the time index $s = 1, \dots, T$, let us define $L_s = \min(Y_{is}| i = 1, \dots, N)$, i.e., $L_s$ is the lowest yield in the $s$-th column of $Y_{is}$. In a normal yield curve this will be the lowest maturity yield (which in our data is 1 Mo). However, there can be situations where this is not the case. Let us now define the de-noised $Z_{is}$ matrix as follows: $Z_{is} = Y_{is} - L_s$. By definition $Z_{is}$ is nonnegative. Also, $L_s$ can be thought of as the level factor (see below) and we can hope that subtracting it from the data will reduce the noise. Basically, $Z_{is}$ is the spread between the yield for a given maturity labeled by $i$ and the lowest yield. We can now apply NMF to this de-noised matrix $Z_{is}$. The results for $K = 2$ are given in Tables 5-6 and Figures 11-14. The results for $K = 3$ are given in Tables 7-8 and Figures 15-20.

For $K = 2$ the errors (both for the weights, which are shown in Table 5, and the factors, which are not shown as the number of dates $T$ is large) are tiny. The overall fit (Table 6) for the de-noised matrix $Z_{is}$ is worse than the vanilla NMF fit (Table 2).[13] However, vanilla NMF has large errors for the weights and factors. In fact, a single vanilla NMF run (i.e., $P = 1$, without any averaging) typically will produce an even better fit than that in Table 2. However, this is meaningless as NMF is nondeterministic and each new run, while superfluously producing an excellent-looking fit, will be sizably different from the one before. In fact, this appears to be a commonly overlooked pitfall when applying NMF.[14] Put another way, the rosy fit in Table 2 is meaningless as the errors in the weights and the factors are too large for the fit to be useful.

Furthermore, for $K = 3$ the errors (for the weights they are shown in Table 7; we do not show the errors for the factors as $T$ is large) are no longer tiny but still smaller than for vanilla NMF. On the other hand, the overall fit for $K = 3$ (Table 8) is better than for $K = 2$ (Table 6).

---

[13] The fit for the original matrix $Y_{is} = Z_{is} + L_s$ is better than that for $Z_{is}$; however, here we are not modeling $L_s$.
[14] This appears to be the case in the analysis of [Takada and Stern, 2015]. Also, one of the motivations behind [Kakushadze and Yu, 2016a] was precisely that this was routinely the case in applying NMF to cancer signatures.

So, while including the third factor superfluously improves the overall fit, it also introduces sizable errors in (some) weights and/or factors. In this regard, it is instructive to look at the *serial* pairwise correlations between the factors $\varphi_{AB} = \text{Cor}(F_{As}, F_{Bs})$ ($A \neq B$) and between the factors and the level $\vartheta_A = \text{Cor}(F_{As}, L_s)$. For $K = 2$ we have: $\varphi_{12} \approx -70.82\%$, $\vartheta_1 \approx 44.6\%$ and $\vartheta_2 \approx -67.77\%$. Importantly, these correlations are stable from averaging over different sets of $P = 100$ runs. This is because for $K = 2$ we have tiny errors, so the local minima that each NMF run finds are very close to the global minimum. However, for $K = 3$ we have substantial variability in the correlations $\varphi_{AB}$ and $\vartheta_A$, which are summarized in Table 9 for 5 different sets of $P = 100$ de-noised NMF runs. We therefore conclude that the "better" overall fit for $K = 3$ is spurious as the third factor introduces substantial instability, so the number of stable factors we can infer from our data is $K = 2$, that is, along with the level $L_s$.

Before we conclude this subsection, let us mention a tweak we can apply to the above analysis. Since we have factored out the level $L_s$, we can remove the shortest maturity (1 Mo) from the dataset altogether and run NMF on the so-reduced data. However, unsurprisingly, this does not alter the results discussed above: $K = 2$ is just as stable with tiny errors, while $K = 3$ has the same instabilities as above. All in all, our conclusions above do appear to hold.

### 2.4. Interpreting the Factors

It is tempting to interpret the $K = 2$ de-noised NMF factors via steepness (a.k.a. slope) and curvature, as was done in [Diebold and Li, 2006] in the context of the 3-factor model of [Nelson and Siegel, 1987], which (in the parametrization of [Diebold and Li, 2006] and conformed to our notations here) is given by

$$Y_{is} \approx \beta_{0s} + \omega_{i1}\,\beta_{1s} + \omega_{i2}\,\beta_{2s} \tag{4}$$

$$\omega_{i1} = \frac{1 - \exp(-\lambda\,\tau_i)}{\lambda\,\tau_i} \tag{5}$$

$$\omega_{i2} = \omega_{i1} - \exp(-\lambda\,\tau_i) \tag{6}$$

Here: $\lambda$ is a parameter (which generally has to be fitted using data, albeit it is fixed differently in [Diebold and Li, 2006]);[15] $\tau_i$ are the maturities; $\omega_{i1}$ and $\omega_{i2}$ are the loadings analogous to our weights $W_{iA}$; $\beta_{1s}$ and $\beta_{2s}$ are analogous to our factors $F_{As}$; and the level $\beta_{0s}$ is analogous to our level $L_s$ except that $\beta_{0s}$ is interpreted as a *long-horizon* factor (as $Y_{is} \approx \beta_{0s}$ for large maturities $\tau_i$), while our $L_s$ typically is a *short-horizon* factor. In [Diebold and Li, 2006] $\beta_{1s}$ and $\beta_{2s}$ are interpreted as the slope and curvature, respectively, with the slope $S_s$ defined as $S_s = Y_{10\text{Yr},s} - Y_{3\text{Mo},s}$ (the 10 Yr yield minus the 3 Mo yield on a given date $s$), and the curvature $C_s$ defined as $C_s = 2\,Y_{2\text{Yr},s} - Y_{10\text{Yr},s} - Y_{3\text{Mo},s}$ (twice the 2 Yr yield minus the 10 Yr yield minus the 3 Mo yield).

---

[15] Moreover, a priori this parameter can depend on the time index $s$; however, this would make it less predictive.

So, here we can ask whether we can interpret our factors $F_{As}$ in terms of the slope $S_s$ and the curvature $C_s$. One "hiccup" here is that, for a normal (upward-sloping) yield curve, $\beta_{0s}$ is a long-horizon factor, while our $L_s$ is a short-horizon factor. On the other hand, if the yield curve is inverted (not only downward-sloping but also "inverted humped" curve, which has occurred lately), using the long-horizon factor $\beta_{0s}$ as the level may well be suboptimal and our definition of the level as the minimum maturity may be more justified. So, a direct comparison of de-noised NMF discussed above with the model of [Nelson and Siegel, 1987] (that is, in the parametrization of [Diebold and Li, 2006]) may not be particularly meaningful or useful. Below we will discuss an alternative definition of the level in the context of de-noised NMF. However, with the above definition of $L_s$ it is still meaningful to inquire if our factors $F_{As}$ might be related to the slope $S_s$ and the curvature $C_s$, irrespective of the model of [Nelson and Siegel, 1987].

First, for the period in our dataset, the serial correlation between the slope $S_s$ and the curvature $C_s$ is high, approximately 90.16%, so only one of these factors can be useful in interpreting our factors $F_{As}$. The first factor $F_{1s}$ (Figure 11) has 92.21% correlation with the slope $S_s$ (and 87.11% correlation with the curvature $C_s$). So, we can interpret the first factor $F_{1s}$ as the slope. The second factor $F_{2s}$ (Figure 12) has a large negative, $-88.34\%$, correlation with the curvature $C_s$ defined as above, to wit, $C_s = 2\ Y_{2\text{Yr},s} - Y_{10\text{Yr},s} - Y_{3\text{Mo},s}$ (and $-87.24\%$ correlation with the slope $S_s$). So, with a grain of salt, we can interpret the second factor $F_{2s}$ as the *negative* curvature $-C_s$. In [Diebold and Li, 2006] $\beta_{2s}$ was interpreted as the curvature $C_s$, which is defined to be positive for a normal curve (which is concave) and negative for an inverted curve (which is convex, at least in some segment).[16] So, it should come as no surprise that for an inverted curve we have $-C_s$ as one of the factors. Once again, this is to be taken with a grain of salt as the slope $S_s$ and the curvature $C_s$ are highly correlated in this dataset.

### 2.5. Alternative De-Noising

Above we de-noised the matrix $Y_{is}$ by subtracting from it, for each date $s$, the lowest yield on that date, which we identify with the level $L_s$. This is a natural thing to do with the view of having a nonnegative de-noised matrix. One consequence of this de-noising is that the level $L_s$ typically (but not always) corresponds to shorter maturities (which tend to be volatile).

There is an alternative way of de-noising the matrix $Y_{is}$. For each value of the time index $s = 1, \ldots, T$, let us define $\tilde{L}_s = \max(Y_{is}|\ i = 1, \ldots, N)$, i.e., $\tilde{L}_s$ is the highest yield in the $s$-th column of $Y_{is}$. In a normal yield curve this will be the longest maturity yield (which in our data is 30 Yr). However, there can be situations where this is not the case. Let us now define the de-noised $\tilde{Z}_{is}$ matrix as follows: $\tilde{Z}_{is} = \tilde{L}_s - Y_{is}$. By definition $\tilde{Z}_{is}$ is nonnegative. Also, $\tilde{L}_s$ can be thought of as the level factor, except that this time it is a long-horizon factor for a normal curve. So, $\tilde{Z}_{is}$ is the spread between the highest yield and the yield for a given maturity labeled by $i$.

---

[16] This is opposite to the standard definition: the curvature is positive (negative) for convex (concave) functions.

We can now apply NMF to this de-noised matrix $\tilde{Z}_{is}$. The results for $K = 2$ are given in Tables 10-11 and Figures 21-24.[17] (The results for $K = 3$, expectedly, are noisy, so are not given.) The results are similar to those in Tables 5-6 except that the correlation $\rho_i$ for the 30 Yr maturity is low. However, this correlation is not meaningful and has no import for the overall fit as the 30 Yr maturity with this alternative de-noising should be dropped altogether. Indeed, out of $T = 276$ dates in the time series, only 12 dates have nonzero $\tilde{Z}_{is}$ for the 30 Yr maturity. This is why the corresponding correlation $\rho_i$ is low, and also why the first weight (W1) in the last row of Table 10 is 0 and the second weight (W2) in the same row is small. So, we can remove the 30 Yr maturity from $\tilde{Z}_{is}$ altogether and run NMF on the so-reduced dataset. The results are only slightly different from those in Tables 10-11 (with the 30 Yr maturity dropped).

For the serial pairwise correlations between the factors $\varphi_{AB} = \text{Cor}(F_{As}, F_{Bs})$ $(A \neq B)$ and between the factors and the level $\vartheta_A = \text{Cor}(F_{As}, \tilde{L}_s)$ (see Subsection 2.3) we have: $\varphi_{12} \approx -87.53\%$, $\vartheta_1 \approx 75\%$ and $\vartheta_2 \approx -79.74\%$. The first factor $F_{1s}$ (Figure 21) has 98.15% correlation with the slope $S_s$ (and 83.95% correlation with the curvature $C_s$). So, we can interpret the first factor $F_{1s}$ as the slope. The second factor $F_{2s}$ (Figure 22) has a large negative, $-96.36\%$, correlation with the curvature $C_s$ defined as above, to wit, $C_s = 2 Y_{2\text{Yr},s} - Y_{10\text{Yr},s} - Y_{3\text{Mo},s}$ (and $-94.25\%$ correlation with the slope $S_s$). So, with a grain of salt (as $S_s$ and $C_s$ have 90.16% correlation), we can also interpret the second factor $F_{2s}$ as the negative curvature $-C_s$.

Using $\tilde{L}_s$ vs. $L_s$ as the level is not a matter of principle but depends on the circumstances such as whether the curve is normal, inverted, (inverted) humped, etc. With NMF we cannot use an arbitrary fixed maturity as the level (e.g., in [Diebold and Li, 2006] the 10 Yr yield was used as the level) for the simple reason that factoring it out from the matrix $Y_{is}$ will produce negative entries thereby defying the purpose as the input data for NMF must be nonnegative.

## 3. Statistical Cluster Factors

A connection between NMF and clustering (and k-means in particular) has been long appreciated (see, e.g., [Ding, He and Simon, 2005], [Zass and Shashua, 2005], [Shahnaz et al, 2006]). In some cases NMF can essentially be clustering in disguise (see, e.g., [Kakushadze and Yu, 2016a] in the context of cancer signatures), to wit, when the weights matrix has a structure where many weights, while nonzero, are relatively small, so there is a semblance of a clustering structure. It is therefore natural to wonder whether there is an underlying clustering structure in the Treasury yields, especially that maturities "naturally" split into short, medium and long. Also, since the number of maturities is small, one may hope to have stability in the clusterings.

---

[17] Note that the weights in Figures 23-24 are "upside-down" compared with the weights in Figures 13-14. This is because in the latter case the de-noised matrix is the spread between the yield and the lowest yield, while in the former case the de-noised matrix is the spread between the yield and the highest yield, hence the "flipping".

Basically, there are two main parts two the story here. If we apply nondeterministic clustering such as k-means to yields (or their normalized versions – see below), we can get different clusterings from different k-means runs, which is conceptually similar to what transpires in NMF. One way of dealing with this is by taking a statistical approach as in [Kakushadze and Yu, 2016b], which was originally developed in the context of equities and later adapted in the context of cancer signatures [Kakushadze and Yu, 2017b, 2017c]. The idea is simple. Let us have $P$ different k-means runs. Each k-means run labeled by $r$ ($r = 1, \ldots, P$) produces a clustering with $K$ clusters, which maps the $N$ vectors labeled by $i$ ($i = 1, \ldots, N$) to $K$ clusters. For a given k-means run labeled by $r$, let us denote the corresponding clustering map via $G^{(r)}$ (so $G^{(r)}: \{1, \ldots, N\} \to \{1, \ldots, K\}$). We can aggregate these clusterings (assuming the clusters in different clusterings are aligned – see below) by adding the $N \times K$ binary matrices $\delta_{G^{(r)}(i), A}$ from the $P$ runs ($G^{(r)}(i)$ labels the cluster to which the vector labeled by $i$ belongs). The so-aggregated $N \times K$ matrix $Q_{iA}$ is not binary: it is a matrix of counts with nonnegative elements. We can now generate a binary matrix $H_{iA}$ from this counts matrix by setting, for each value of $i$, $H_{iA} = 1$ for the value of $A$ for which $Q_{iA}$ is the largest; otherwise $H_{iA} = 0$. A priori there can be ties in this process (i.e., for a given value of $i$, there can be more than one elements in $\text{argmax}_B Q_{iB}$), which can be resolved by taking the most populous cluster among the ties (see [Kakushadze and Yu, 2016b] for details). The binary matrix $H_{iA}$ defines a binary clustering map $G: \{1, \ldots, N\} \to \{1, \ldots, K\}$, where $H_{iA} = \delta_{G(i), A}$. So, this way we can remove the nondeterminism of k-means subject to the following two "caveats" (which will be resolved).

First, above we assume that the clusters in different clusterings corresponding to the $P$ different k-means runs are aligned. However, just as the factors in different NMF runs are not necessarily aligned, here too we have no guarantee that the clusters are aligned. In fact, these clusters can look rather different from run to run. However, just as in NMF, we can bootstrap the cluster centers (which are $K \times T$ matrices) row-wise, obtain a $(P \cdot K) \times T$ matrix this way, and then cluster its $(P \cdot K)$ rows into $K$ clusters via k-means, thereby aligning the clusters from the $P$ different k-means runs (similarly to what we did with NMF). One "hiccup" here is that we may end up with fewer than $K$ clusters. However, this is not problematic; in fact, it means that our original guess for the number of clusters overshoots, and we can simply proceed with the smaller number $K'$ of the resultant clusters. One remaining cloud in the sky is that we use k-means, a nondeterministic algorithm, for aligning the clusters, so the final result may well be nondeterministic, i.e., for different sets of $P$ k-means runs we may end up with different clusterings (albeit this possible remaining nondeterminism can be expected to be much milder than that of k-means, that is, assuming the data indeed has a reasonable underlying clustering structure). This is dealt with via the so-called *K-means algorithm of [Kakushadze and Yu, 2017b]. The idea is simple. We can take a large number $M$ of different sets of $P$ k-means runs (each set gives a clustering) and take the clustering that arises most frequently in these $M$ sets.

*K-means is a *statistically deterministic* algorithm and produces a unique answer. If the size of the data is too large, then the number of k-means runs ($M \cdot P$) can be too large to make sense computationally. However, for smaller datasets, such as the yields data we are working with, this typically is not expected to pose an issue. In fact, in our runs with $K = 2$ and $K = 3$ (see below) on the same data as we used with NMF above, $M = 100$ sets with $P = 100$ runs in each set all produced identical clusterings (so *K-means was not even required for this dataset). This implies that the clustering structure (at least for these values of $K$) is fully stable.

Now that we have discussed how to cluster, we must also address what to cluster. We could cluster the $N$ rows of the matrix $Y_{is}$ (each row being a $T$-vector, and in k-means we can use the default Euclidean distance between two $T$-vectors as the similarity criterion). However, different maturities (to which the rows of $Y_{is}$ correspond) have different serial volatilities (the shorter maturity yields tend to be more volatile). So, as is common in clustering, we can normalize the rows of $Y_{is}$ as follows: $X_{is} = Y_{is} / \sigma_i$, where $\sigma_i = \sqrt{\text{Var}(Y_{is})}$ are serial volatilities (Var($\cdot$) is a serial variance). We can then cluster $X_{is}$, whose rows are now properly normalized.

The second part of the story is how to obtain the weights once we get the clusters. A simple solution is that we use *one-factor* NMF within each cluster, so we have $K$ factors, all of which, along with all the weights, are nonnegative. One-factor NMF is straightforward to compute. This is because one-factor NMF is equivalent to a singular value decomposition (SVD) truncated to the first eigenvalue. Thus, according to the Eckart-Young-Mirsky theorem [Eckart and Young, 1936], the closest (w.r.t. minimizing the Frobenius norm) rank-$k$ approximation of a matrix is given by the rank-$k$ SVD truncation of said matrix. So, the best rank-1 approximation for an $n \times m$ matrix $A_{is}$ is given by the rank-1 SVD truncation: $A_{is} \approx \sqrt{\lambda}\, u_i\, v_s$, where $u_i$ is the first principal component of the matrix $AA^T$, $v_s$ is the first principal component of $A^T A$, and $\lambda$ is the corresponding eigenvalue (which is the same for both $u_i$ and $v_s$). Next, according to the Perron-Frobenius theorem [Perron, 1907], [Frobenius, 1912], all the elements of the first principal component of a positive matrix are all positive (or can be chosen to be such as the signs thereof can always be flipped simultaneously), so both $u_i$ and $v_s$ are positive if $A_{is}$ is positive, and thus the rank-1 SVD truncation of a positive matrix produces its one-factor NMF.[18]

So, assuming we have $K$ binary clusters $C_A$ ($A = 1, \dots, K$), for a given cluster labeled by $A$, we can compute the weights $W_{iA}$ for $i \in C_A$ by simply taking the first principal component of the matrix $(YY^T)_{ij}$, $i, j \in C_A$, and normalizing it such that its elements add up to 1. We can compute the factors $F_{As}$ by taking the first principal component of the $T \times T$ matrix $\sum_{i \in C_A} Y_{is} Y_{it}$ ($s, t = 1, \dots, T$) and normalizing it accordingly (see above). Also note that $W_{iA} = 0$ for $i \notin C_A$.

---

[18] Interestingly, the R package "NMF" [Gaujoux and Seoighe, 2010] (see Appendix A) for one-factor NMF produces slightly (but not negligibly) worse results than the rank-1 SVD truncation. The R function `foo.nmf(n, m)` in our Appendix C compares one-factor NMF (using said package) with the rank-1 SVD truncation of an $n \times m$ matrix.

The R source code is given in Appendix B. The results for $K = 2$ are given in Tables 13-14[19] and Figures 25-26, and the results for $K = 3$ are given in Tables 15-16 and Figures 27-29.[20] In this clustering-based approach there is no need to de-noise the matrix $Y_{is}$ (or $X_{is}$) as there are no error bars: the result is (statistically) deterministic as we get a unique clustering, and then the weights (and factors) within each cluster are also uniquely determined (the rank-1 SVD truncation corresponds to the global optimum – see above). However, along with the slope $S_s$ and the curvature $C_s$ (see above), as in [Diebold and Li, 2006], it is instructive to also define the level $L_s = Y_{10\text{Yr},s}$ as the 10 Yr yield and compute the correlations for the factors $F_{As}$ with $L_s$, $S_s$ and $C_s$. The latter are themselves highly correlated (serially). Thus: $\text{Cor}(L_s, S_s) \approx 84.57\%$, $\text{Cor}(L_s, C_s) \approx 74.60\%$, and (as already mentioned above) we have $\text{Cor}(S_s, C_s) \approx 90.16\%$.

Now, for $K = 2$ we have the following serial correlations: $\text{Cor}(F_{1s}, L_s) \approx 78.77\%$, $\text{Cor}(F_{1s}, S_s) \approx 33.92\%$, $\text{Cor}(F_{1s}, C_s) \approx 28.67\%$; $\text{Cor}(F_{2s}, L_s) \approx 99.87\%$, $\text{Cor}(F_{2s}, S_s) \approx 82.79\%$, $\text{Cor}(F_{2s}, C_s) \approx 74.54\%$. So, the first factor (Figure 25), which is a short-horizon factor (Table 13), has a relatively high correlation with the level, but low correlations with the slope and the curvature. The second factor (Figure 26), which contains medium and long maturities (Table 13), is almost 100% correlated with the level and has relatively high correlations with the slope and the curvature. The serial correlation between the factors $\varphi_{12} = \text{Cor}(F_{1s}, F_{2s}) \approx 80.69\%$ is high. The fit (Table 14) is good (and better than for NMF, even without de-noising).

Next, for $K = 3$ we have the following serial correlations: $\text{Cor}(F_{1s}, L_s) \approx 73.89\%$, $\text{Cor}(F_{1s}, S_s) \approx 26.75\%$, $\text{Cor}(F_{1s}, C_s) \approx 21.54\%$; $\text{Cor}(F_{2s}, L_s) \approx 97.29\%$, $\text{Cor}(F_{2s}, S_s) \approx 70.26\%$, $\text{Cor}(F_{2s}, C_s) \approx 62.21\%$; $\text{Cor}(F_{3s}, L_s) \approx 99.89\%$, $\text{Cor}(F_{3s}, S_s) \approx 84.52\%$, $\text{Cor}(F_{3s}, C_s) \approx 76.51\%$. So, the first factor (Figure 27), which is a short-horizon factor (Table 15), has a relatively high correlation with the level, but low correlations with the slope and the curvature (similarly to the $K = 2$ case). The second factor (Figure 28) is built from the 6 Mo, 1 Yr and (perhaps surprisingly) 30 Yr maturities (Table 15). The third factor (Figure 29) is built from medium and long maturities (Table 15). Both the second and the third factors are almost 100% correlated with the level and have relatively high correlations with the slope and the curvature (especially the third factor). The serial correlations between the first and the other two factors are relatively high, and the correlation between the second and the third factors is very high: $\varphi_{12} \approx 87.10\%$, $\varphi_{13} \approx 73.77\%$, $\varphi_{23} \approx 97.26\%$. The overall fit (Table 14) is good. However, $K = 3$ does not appear to add value compared with $K = 2$: the second and the third factors are too highly correlated and have similar correlations with the level, the slope and the curvature. This can be traced to the fact that the level, the slope and the curvature are already highly correlated. The eRank (see above) of their correlation matrix is 1.51. So, it appears to make sense to model the yields with a 2-factor model (at least for this time period) instead of 3.

---

[19] Table 12, which we refer to in Appendix A, gives some sample Treasury yield data from our dataset (see above).
[20] We do not plot the weights as they vanish outside the clusters, while the within-cluster weights are rather close.

### *3.1. What About Forecasting?*

Forecasting Treasury yields based on historical yield data alone generally is challenging (see, e.g., [Duffee, 2002], [Diebold and Li, 2006], [Duffee, 2013], [Almeida et al, 2018], and references therein). Using machine learning methods such as NMF and clustering discussed above can achieve rosy-looking fits in-sample. However, out-of-sample forecasting is still not a cake walk by any stretch. There are basically two parts to the story here, to which we now turn.

So, in the factor model context, where we approximate the matrix $Y \approx W\,F$, we have the weights $W_{iA}$ and the factors $F_{As}$. To be able to forecast $Y_{is}$ out-of-sample, we must be able to forecast the factors $F_{As}$. Looking even at the *in-sample* plots of the factors, e.g., Figures 25-26, it is clear that the factors have sizable stochastic (noise) components to them (which we can expect to get even worse out-of-sample), and they further have nontrivial temporal dynamics with regime changes, etc. So, it is clear that forecasting the factors $F_{As}$ is challenging. E.g., we can try to identify (short-horizon) trends ("momentum") in the factors and forecast changes based on such trends. However, because of the inherent noise, such "momentum" (computed as, e.g., the temporal slope in a given factor) will have sizable errors thereby affecting the forecasted value of said factor. Furthermore, such forecasting produces notoriously poor results when trend reversals take place, and not much can be done with that.[21] *Así es la vida*.

However, even if somehow – magically – we could forecast the factors with high accuracy (which is challenging), this by itself would not suffice for accurately forecasting the yields. This is because we also must worry about the weights $W_{iA}$, which can and do sizably change from period to period thereby resulting in *out-of-sample* instability. This is a typical pitfall of using machine learning methods that attempt to learn almost everything (barring hyperparameters such as the number of factors/clusters $K$) from the data itself. In this case this amounts to determining the weights $W_{iA}$ (which are the loadings matrices in the regression nomenclature) from the data as opposed to fixing them based on some fundamental or holistic considerations. However, the price one must pay is that these weights are nonstationary. This can be seen from Figures 30-31, where we plot the weights $W_{iA}$ in the $K = 2$ cluster model above computed using 21-trading-day (that is, monthly) periods into which we break our data. In Figures 32-33 we plot the weights $W_{iA}$ in the $K = 2$ cluster model above computed daily (i.e., $\sum_{A=1}^{K} W_{iA}\,F_{As}$ is fitted into $Y_{is}$ for each date $s$ separately to compute the corresponding weights $W_{iA}$ daily). Unsurprisingly, the daily weights (Figures 32-33) are quite noisy. Both the 21-day and daily weights are nonstationary, which makes forecasting challenging. As an aside, note that the short-maturity weights (Figures 30 and 32) have been converging (flattening segment), while the long-maturity weights (Figures 30 and 32) have been diverging (steepening segment).

---

[21] In this regard, autoregressive models such as AR(1) (which was utilized in [Diebold and Li, 2006] for forecasting the factors in the context of the model of [Nelson and Siegel, 1987]) and similar approaches may be of little help.

## 4. Concluding Remarks

In light of our discussion above on forecasting, it is natural to wonder if and how the forecasting challenges are different in other factor model approaches. For instance, factor models based on (the first three) principal components (see [Litterman and Scheinkman, 1991], [Bliss, 1997]) have been studied in detail.[22] However, forecasting in such models is also challenging. This is because higher-than-first principal components are notoriously unstable out-of-sample. The first principal component tends to be less unstable. However, weights based on the first principal component are still substantially nonstationary. Thus, note that in the clustering models we discussed in Section 3, the weights are the within-cluster first principal components.[23] And they are significantly unstable out-of-sample. So, to recap, higher principal components make out-of-sample instability even worse. Also, on another note, some weights (and factors) in the principal component approach are negative and lack the (at least superfluously) appealing nonnegativity property of NMF and the models based on clustering.

From the forecasting viewpoint, the approach of [Diebold and Li, 2006] in using the parametric model of [Nelson and Siegel, 1987][24] (see Eqns. (4), (5) and (6)) would appear to be appealing. In this model the weights $W_{iA}$ parametrically depend on the maturities $\tau_i$, through the sole parameter $\lambda$ (which in [Diebold and Li, 2006] was not fitted but fixed using exogenous considerations), which might appear to bode well with out-of-sample stability. However, this parameter $\lambda$ itself is nonstationary and varies from period to period. Furthermore, while the weights in this model are expressly positive, for this model to explain non-upward-sloping curves (downwards-sloping, "humped", "inverted humped"), some factors must be negative. And at least for shorter horizons (e.g., 1 month) the model does not forecast well [Diebold and Li, 2006]. So, while a parametric approach has its appeals, apparently, there is no free lunch.

Let us also mention that, while above we discussed forecasting in the context of the clustering models of Section 3, the same conclusions apply to NMF-based models. There too, as in any similar machine learning method, the weights are learned from the data and there is no reason why they would be stationary, and typically in such problems they are far from it.

## Appendix A: R Source Code for NMF Algorithm

In this Appendix we give R (R Project for Statistical Computing, https://www.r-project.org/) source code for the vanilla and de-noised NMF algorithms discussed in Section 2. The source code consist of a single function `treasury(k, n = 100, denoise)` and is straightforward. Internally it loads and uses the R package "NMF" [Gaujoux and Seoighe, 2010]. The inputs of `treasury()` are: the number of factors `k` to try (which is denoted by $K$ in the main text); the number `n` of NMF runs to average over (which is denoted by $P$ in the main text); and `denoise`, whose values are as follows: `denoise = 0` corresponds to vanilla NMF (without de-noising); `denoise = 1` corresponds to de-noised NMF discussed in Subsection 2.3; and `denoise = 2` corresponds to NMF with alternative de-noising discussed in Subsection 2.5. Furthermore, the `treasury()` function internally reads a data file (which is a flat tab-delimited text file) `treasury.txt`. This file is a $(T + 1) \times (N + 1)$ table containing the Treasury yields data, which can be freely downloaded from https://www.treasury.gov/resource-center/data-chart-center/interest-rates/pages/TextView.aspx?data=yieldAll. Its first row is the column labels. Below the first row, the first column is the $T$ dates, and the other 12 columns are the Treasury yields corresponding to the $N = 12$ maturities (1 Mo, 2 Mo, 3 Mo, 6 Mo, 1 Yr, 2 Yr, 3 Yr, 5 Yr, 7 Yr, 10 Yr, 20 Yr and 30 Yr). To aid with visualizing this data, sample data from the `treasury.txt` file is given in Table 12. The function `treasury()` outputs two (flat tab-delimited) text files, one with the weights $W_{iA}$ (averaged over `n` NMF runs) and the corresponding standard deviations, and the other with the correlations $\rho_i$ and errors $E_i$ (see Subsection 2.2 for details). It also outputs JPEG files with the plots of the weights $W_{iA}$ (vs. maturity) and factors $F_{As}$ (vs. time). Finally, it prints on-screen various quantities discussed in Section 2, such as the effective rank (eRank) of the serial correlation matrix $\Psi_{ij}$ between different maturities, the serial correlation matrix between the factors (which include the level when `denoise = 1` or `denoise = 2`), and the serial correlation matrix between the factors (which include the level when `denoise = 1` or `denoise = 2`) and the slope $S_s$ and the curvature $C_s$. Internally the function `treasury()` calls a subfunction `calc.erank()`, which is a subfunction of the function `qrm.erank.pc()` in Appendix A of [Kakushadze and Yu, 2017a].

---

[22] Including an application of this approach to short-maturity instruments [Knez, Litterman and Scheinkman, 1994].
[23] Thus, the clustering models of Section 3 are conceptually similar to the equity risk models of [Kakushadze, 2015].
[24] Note that in our analyses on our dataset we did not find the "second curvature" (4th) factor of [Svensson, 1994]. However, this paper is not intended to be empirically exhaustive and other time periods may turn up other factors.

```
treasury <- function (k, n = 100, denoise)
{
    require(NMF)

    no.na <- function(x)
    {
        return(!any(x == "N/A"))
    }

    calc.erank <- function(x, excl.first)
    {
        take <- x > 0
        x <- x[take]
        if(excl.first)
            x <- x[-1]
        p <- x / sum(x)
        h <- - sum(p * log(p))
        er <- exp(h)
```

```r
        if(excl.first)
            er <- er + 1
        return(er)
}

x <- read.delim("treasury.txt", header = F)
x <- as.matrix(x)
hdr <- x[1, ]
hdr <- hdr[-1]
x <- x[-1, ]
d <- x[, 1]
x <- x[, -1]
take <- apply(x, 1, no.na)
x <- x[take, ]
mode(x) <- "numeric"

q <- cor(x)
q1 <- q
diag(q1) <- NA
print(paste("Average pairwise correlation = ",
      round(mean(q1, na.rm = T) * 100, 2), sep = ""))
p <- eigen(q)$values
print(paste("eRank = ",
      round(calc.erank(p, excl.first = F), 2), sep = ""))
print(paste("ModeRank = ",
      round(calc.erank(p, excl.first = T), 2), sep = ""))

slope <- x[, 10] - x[, 3]
curv <- 2 * x[, 6] - x[, 10] - x[, 3]

if(denoise == 1)
{
     lvl <- apply(x, 1, min)
     x <- x - lvl
     ### x <- x[, -1] # Remove 1 Mo maturity
     ### hdr <- hdr[-1]
}

if(denoise == 2)
{
     lvl <- apply(x, 1, max)
     x <- lvl - x
     ### x <- x[, -12] # Remove 30 Yr maturity
     ### hdr <- hdr[-12]
}

x <- t(x)
red.k <- T
while(red.k)
{
     print(paste("Trying k = ", k, sep = ""))
     w.av <- w.sd <- w.med <- w.mad <- matrix(0, nrow(x), k)
```

```r
y.av <- y.sd <- y.med <- y.mad <- matrix(0, k, ncol(x))
w.b <- matrix(0, nrow(x), k * n)
y.b <- matrix(0, k * n, ncol(x))
for(i in 1:n)
{
    v <- nmf(x, rank = k, nrun = 1)
    w <- basis(v)
    w.n <- colSums(w)
    w <- t(t(w) / w.n)
    y <- coef(v)
    y <- y * w.n
    w.b[, (1:k) + (i - 1) * k] <- w
    y.b[(1:k) + (i - 1) * k, ] <- y
}
if(n > 1)
{
    cl.w <- kmeans(t(w.b), k, iter.max = 100)
    cl.w <- cl.w$cluster
}
else
{
    w.av <- w.med <- w.b
    w.sd[] <- w.mad[] <- 0
    y.av <- y.med <- y.b
    y.sd[] <- y.mad[] <- 0
    break
}
red.k <- F
for(j in 1:k)
{
    take.w <- cl.w == j
    print(paste("Number of elements in cluster ",
        j, " = ", sum(take.w), sep = ""))
    if(sum(take.w) > 1)
    {
        w.av[, j] <- rowMeans(w.b[, take.w])
        w.sd[, j] <- apply(w.b[, take.w], 1, sd)
        w.med[, j] <- apply(w.b[, take.w], 1, median)
        w.mad[, j] <- apply(w.b[, take.w], 1, mad)
        y.av[j, ] <- colMeans(y.b[take.w, ])
        y.sd[j, ] <- apply(y.b[take.w, ], 2, sd)
        y.med[j, ] <- apply(y.b[take.w, ], 2, median)
        y.mad[j, ] <- apply(y.b[take.w, ], 2, mad)
    }
    else
    {
        print("Reducing k")
        red.k <- T
        k <- k - 1
        break
    }
}
```

```r
}

time.stamp <- paste(Sys.Date(), ".",
      format(Sys.time(), "%H%M%S"), sep = "")
days <- ncol(y.av)
my.col <- c("green", "red", "blue", "black")
for(j in 1:nrow(y.av))
{
      file <- paste("Factor.", j, ".", time.stamp,
            ".jpeg", sep = "")
      jpeg(file = file, width = 1800, height = 1800,
            units = "px", res = 300)
      col <- my.col[j]
      y.max <- max(y.av[j, ] + 1.1 * y.sd[j, ])
      y.min <- min(y.av[j, ] - 1.1 * y.sd[j, ])
      plot(1:days, y.av[j, ], type = "l",
            col = col, xlab = "Days", ylab = "Factor",
            ylim = c(y.min, y.max))
      lines(1:days, y.av[j, ] + y.sd[j, ], col = col, lty = 3)
      lines(1:days, y.av[j, ] - y.sd[j, ], col = col, lty = 3)
      dev.off()

      mat <- log(c(1,2,3,6,12,24,36,60,84,120,240,360))
      ### mat <- mat[-1] # Remove 1 Mo maturity
      ### mat <- mat[-12] # Remove 30 Yr maturity

      file <- paste("Weights.", j, ".", time.stamp,
            ".jpeg", sep = "")
      jpeg(file = file, width = 1800, height = 1800,
            units = "px", res = 300)
      col <- my.col[j]
      w.max <- max(w.av[, j] + 1.1 * w.sd[, j])
      w.min <- min(w.av[, j] - 1.1 * w.sd[, j])
      plot(mat, w.av[, j], type = "l",
            col = col, xlab = "Log(Maturity)",
            ylab = "Weight", ylim = c(w.min, w.max))
      lines(mat, w.av[, j] + w.sd[, j], col = col, lty = 3)
      lines(mat, w.av[, j] - w.sd[, j], col = col, lty = 3)
      dev.off()
}

x.fit <- w.av %*% y.av
if(denoise)
      w <- cbind(round(w.av * 100, 2), round(w.sd * 100, 6))
else
      w <- round(cbind(w.av, w.sd) * 100, 2)

w <- cbind(hdr, w)
file <- paste("w.", k, ".", n, ".", time.stamp, ".txt", sep = "")
write.table(w, file = file, quote = F,
      row.names = F, col.names = F, sep = "\t")
r <- ss <- rep(NA, nrow(x))
```

```
        for(j in 1:nrow(x))
        {
            r[j] <- cor(x[j, ], x.fit[j, ])
            ss[j] <- sum((x[j, ] - x.fit[j, ])^2)
        }
        rss <- round(cbind(r * 100, ss), 2)
        rss <- cbind(hdr, rss)
        file <- paste("rss.", k, ".", n, ".", time.stamp,
            ".txt", sep = "")
        write.table(rss, file = file, quote = F,
            row.names = F, col.names = F, sep = "\t")

        if(denoise)
            fac <- t(rbind(lvl, y.av))
        else
            fac <- t(y.av)

        print(paste("Correlation between slope and curvature = ",
            round(cor(slope, curv) * 100, 2), sep = ""))
        print("Factor correlation matrix:")
        print(round(cor(fac) * 100, 2))
        print("Correlation matrix between factors & slope + curvature:")
        print(round(cor(fac, cbind(slope, curv)) * 100, 2))
}
```

## Appendix B: R Source Code for Clustering Algorithm

In this Appendix we give R source code for the clustering-based algorithms of Section 3. The source code consist of a single function `treasury.cl(k, n = 100)` and is straightforward. Internally `treasury.cl()` calls the R function `qrm.stat.ind.class()`, which is given in Appendix A of [Kakushadze and Yu, 2016b]. This function in turn internally calls other functions, which are also given in Appendix A of [Kakushadze and Yu, 2016b]. However, one of those functions, to wit, `qrm.calc.norm.ret()`, is redefined in `treasury.cl()` as this function normalizes the quantities to be clustered, and this normalization is different in the context of equities (which is the focus of [Kakushadze and Yu, 2016b]) and in the context of Treasury yields (for which we discuss the normalization in Section 3). The inputs of `treasury.cl()` are: the number of clusters `k` (which is denoted by $K$ in the main text); and the number `n` of the k-means runs to aggregate (which is passed into `qrm.stat.ind.class()` and is denoted by $P$ in the main text), which we also use as the number of sets $M$ of $P$ k-means runs (see Section 3), i.e., we set $M = P$ (even though these two parameters are independent). The `treasury.cl()` function internally reads a tab-delimited text file `treasury.txt`, which is described in Appendix A. The `treasury.cl()` function outputs two text files, one with the weights $W_{iA}$ within each cluster, and the other with the correlations $\rho_i$ and errors $E_i$ (see Subsection 2.2 for details). It also outputs JPEG files with the plots of the factors $F_{As}$ (these are serial plots). Finally, it prints on-screen various quantities discussed in Section 3, such as the serial correlation matrix between the factors, and the serial correlations between the factors and the level, the slope and the curvature (see Section 3).

```r
treasury.cl <- function (k, n = 100)
{
      qrm.calc.norm.ret <- function(x)
      {
          return(x / apply(x, 1, sd))
      }

      no.na <- function(x)
      {
          return(!any(x == "N/A"))
      }

      x <- read.delim("treasury.txt", header = F)
      x <- as.matrix(x)
      hdr <- x[1, ]
      hdr <- hdr[-1]
      x <- x[-1, ]
      d <- x[, 1]
      x <- x[, -1]
      take <- apply(x, 1, no.na)
      x <- x[take, ]
      mode(x) <- "numeric"

      lvl <- x[, 10]
      slope <- x[, 10] - x[, 3]
      curv <- 2 * x[, 6] - x[, 10] - x[, 3]

      x <- t(x)
      p <- matrix(NA, n, nrow(x))
      for(j in 1:n)
      {
          z <- qrm.stat.ind.class(x, k, iter.max = 100, num.try = n)
          y <- residuals(lm(x ~ z))
          p[j, ] <- rowSums(y^2)
      }
      p <- t(t(p) - p[1, ])
      p <- round(p, 10)
      if(sum(p > 0) > 0)
          stop("Unstable clustering, use *K-means.")

      w <- matrix(0, nrow(x), k)
      y <- matrix(0, k, ncol(x))
      for(j in 1:ncol(z))
      {
          take <- z[, j] == 1
          q <- eigen(x[take, ] %*% t(x[take, ]))
          w[take, j] <- sqrt(q$values[1]) * abs(q$vectors[, 1])
          q <- eigen(t(x[take, ]) %*% x[take, ])
          y[j, ] <- abs(q$vectors[, 1])
      }
      w.n <- colSums(w)
      w <- t(t(w) / w.n)
```

```r
    y <- y * w.n

    time.stamp <- paste(Sys.Date(), ".",
        format(Sys.time(), "%H%M%S"), sep = "")
    days <- ncol(y)
    my.col <- c("green", "red", "blue", "black")
    for(j in 1:nrow(y))
    {
        file <- paste("Factor.", j, ".", time.stamp, ".jpeg",
            sep = "")
        jpeg(file = file, width = 1800, height = 1800,
            units = "px", res = 300)
        col <- my.col[j]
        plot(1:days, y[j, ], type = "l", col = col,
            xlab = "Days", ylab = "Factor")
        dev.off()
    }

    x.fit <- w %*% y
    w <- cbind(hdr, round(w * 100, 2))
    file <- paste("w.", k, ".", n, ".", time.stamp, ".txt", sep = "")
    write.table(w, file = file, quote = F,
        row.names = F, col.names = F, sep = "\t")
    r <- ss <- rep(NA, nrow(x))
    for(j in 1:nrow(x))
    {
        r[j] <- cor(x[j, ], x.fit[j, ])
        ss[j] <- sum((x[j, ] - x.fit[j, ])^2)
    }
    rss <- round(cbind(r * 100, ss), 2)
    rss <- cbind(hdr, rss)
    file <- paste("rss.", k, ".", n, ".", time.stamp, ".txt",
        sep = "")
    write.table(rss, file = file, quote = F,
        row.names = F, col.names = F, sep = "\t")
    fac <- t(y)
    print("Correlations between level, slope, curvature:")
    print(round(cor(cbind(lvl, slope, curv)) * 100, 2))
    print("Factor correlation matrix:")
    print(round(cor(fac) * 100, 2))
    print("Correlations between factors & level, slope, curvature:")
    print(round(cor(fac, cbind(lvl, slope, curv)) * 100, 2))
}
```

## Appendix C: R Code for One-Factor NMF vs. Rank-1 SVD Comparison

The R function `foo.nmf(n, m)` in this appendix runs one-factor NMF using the R package "NMF" [Gaujoux and Seoighe, 2010] (see Appendix A) vs. the rank-1 SVD truncation of a randomly generated $n \times m$ matrix. As mentioned in the main text, one-factor NMF using said package produces slightly (but not negligibly) worse results than the rank-1 SVD truncation.

```
foo.nmf <- function (n = 10, m = 20)
{
    require(NMF)
    x <- matrix(abs(rnorm(n * m, 0, 1)), n, m)
    y <- nmf(x, rank = 1, nrun = 1)
    x1 <- as.vector(basis(y)) %*% t(as.vector(coef(y)))
    q <- eigen(x %*% t(x))
    x2 <- sqrt(q$values[1]) * abs(q$vectors[, 1])
    q <- abs(eigen(t(x) %*% x)$vectors[, 1])
    x2 <- x2 %*% t(q)
    print(sum((x - x1)^2))
    print(sum((x - x2)^2))
}
```

**Appendix D: DISCLAIMERS**

Wherever the context so requires, the masculine gender includes the feminine and/or neuter, and the singular form includes the plural and vice-versa. The author of this paper ("Author") and his affiliates including without limitation Quantigic® Solutions LLC ("Author's Affiliates" or "his Affiliates") make no implied or express warranties or any other representations whatsoever, including without limitation implied warranties of merchantability and fitness for a particular purpose, in connection with or with regard to the content of this paper including without limitation any code or algorithms contained herein ("Content").

The reader may use the Content solely at his/her/its own risk and the reader shall have no claims whatsoever against the Author or his Affiliates and the Author and his Affiliates shall have no liability whatsoever to the reader or any third party whatsoever for any loss, expense, opportunity cost, damages or any other adverse effects whatsoever relating to or arising from the use of the Content by the reader including without any limitation whatsoever: any direct, indirect, incidental, special, consequential or any other damages incurred by the reader, however caused and under any theory of liability; any loss of profit (whether incurred directly or indirectly), any loss of goodwill or reputation, any loss of data suffered, cost of procurement of substitute goods or services, or any other tangible or intangible loss; any reliance placed by the reader on the completeness, accuracy or existence of the Content or any other effect of using the Content; and any and all other adversities or negative effects the reader might encounter in using the Content irrespective of whether the Author or his Affiliates is or are or should have been aware of such adversities or negative effects.

The R code included in Appendix A, Appendix B and Appendix C hereof is part of the copyrighted R code of Quantigic® Solutions LLC and is provided herein with the express permission of Quantigic® Solutions LLC. The copyright owner retains all rights, title and interest in and to its copyrighted source code included in Appendix A, Appendix B and Appendix C hereof and any and all copyrights therefor.

# References

Almeida, C., Ardison, K., Kubudi, D., Simonsen, A. and Vicente, J. (2018) Forecasting Bond Yields with Segmented Term Structure Models. *Journal of Financial Econometrics* **16**(1): 1-33.

Bliss, R.R. (1997) Movements in the Term Structure of Interest Rates. *Federal Reserve Bank of Atlanta Economic Review* **82**(4): 16-33.

Bouchaud, J.-P. and Potters, M. (2011) Financial applications of random matrix theory: a short review. In: Akemann, G., Baik, J. and Di Francesco, P. (eds.) *The Oxford Handbook of Random Matrix Theory.* Oxford, United Kingdom: Oxford University Press.

Campbell, L.L. (1960) Minimum coefficient rate for stationary random processes. *Information and Control* **3**(4): 360-371.

Diebold, F.X. and Li, C. (2006) Forecasting the Term Structure of Government Bond Yields. *Journal of Econometrics* **130**(2): 337-364.

Ding, C., He, X. and Simon, H.D. (2005) On the equivalence of nonnegative matrix factorization and spectral clustering. In: Kargupta, H., Srivastava, J., Kamath, C. and Goodman, A. (eds.) *Proceedings of the Fifth SIAM International Conference on Data Mining*. Philadelphia, PA: Society for Industrial and Applied Mathematics (SIAM), pp. 606-610.

Duffee, G. (2002) Term premia and interest rate forecasts in affine models. *Journal of Finance* **57**(1): 405-443.

Duffee, G. (2013) Chapter 7 – Forecasting Interest Rates. In: Elliott, G. and Timmermann, A. (eds.) *Handbook of Economic Forecasting*. Vol. 2, Part A. Amsterdam, The Netherlands: Elsevier, pp. 385-426.

Eckart, C. and Young, G. (1936) The approximation of one matrix by another of lower rank. *Psychometrika* **1**(3): 211-218.

Fama, E.F. and MacBeth, J.D. (1973) Risk, Return and Equilibrium: Empirical Tests. *Journal of Political Economy* **81**(3): 607-636.

Forgy, E.W. (1965) Cluster analysis of multivariate data: efficiency versus interpretability of classifications. *Biometrics* **21**(3): 768-769.

Frobenius, G. (1912) Über Matrizen aus Nicht Negativen Elementen. In: *Sitzungsberichte der Königlich Preussischen Akademie der Wissenschaften zu Berlin*, pp. 456-477.

Gaujoux, R. and Seoighe, C. (2010). A flexible R package for nonnegative matrix factorization. *BMC Bioinformatics* **11**: 367.

Hartigan, J.A. (1975) *Clustering Algorithms*. New York, NY: John Wiley & Sons, Inc.

Hartigan, J.A. and Wong, M.A. (1979) Algorithm AS 136: A K-Means Clustering Algorithm. *Journal of the Royal Statistical Society, Series C (Applied Statistics)* **28**(1): 100-108.

Kakushadze, Z. (2015) Heterotic Risk Models. *Wilmott Magazine* **2015**(80): 40-55. Available online: https://ssrn.com/abstract=2600798.

Kakushadze, Z. and Yu, W. (2016a) Factor Models for Cancer Signatures. *Physica A* **462**: 527-559. Available online: https://ssrn.com/abstract=2772458.

Kakushadze, Z. and Yu, W. (2016b) Statistical Industry Classification. *Journal of Risk & Control* **3**(1): 17-65. Available online: https://ssrn.com/abstract=2802753.

Kakushadze, Z. and Yu, W. (2017a) Statistical Risk Models. *Journal of Investment Strategies* **6**(2): 1-40. Available online: https://ssrn.com/abstract=2732453.

Kakushadze, Z. and Yu, W. (2017b) *K-means and Cluster Models for Cancer Signatures. *Biomolecular Detection and Quantification* **13**: 7-31. Available online: https://ssrn.com/abstract=2908286.

Kakushadze, Z. and Yu, W. (2017c) Mutation Clusters from Cancer Exome. *Genes* **8**(8): 201. Available online: https://ssrn.com/abstract=2945010.

Knez, P., Litterman, R.B. and Scheinkman, J. (1994) Explorations into Factors Explaining Money Market Returns. *Journal of Finance* **49**(5): 1861-1882.

Lee, D.D. and Seung, H.S. (1999) Learning the parts of objects by non-negative matrix factorization. *Nature* **401**(6755): 788-791.

Litterman, R.B. and Scheinkman, J. (1991) Common factors affecting bond returns. *Journal of Fixed Income* **1**(1): 54-61.

Lloyd, S.P. (1957) Least square quantization in PCM. *Working Paper.* Murray Hill, NJ: Bell Telephone Laboratories.

Lloyd, S.P. (1982) Least square quantization in PCM. *IEEE Transactions on Information Theory* **28**(2): 129-137.

MacQueen, J.B. (1967) Some Methods for classification and Analysis of Multivariate Observations. In: LeCam, L. and Neyman, J. (eds.) *Proceedings of the 5th Berkeley Symposium* *on Mathematical Statistics and Probability.* Berkeley, CA: University of California Press, pp. 281-297.

Nelson, C. and Siegel, A.F. (1987) Parsimonious modeling of yield curves. *Journal of Business* **60**(4): 473-489.

Paatero, P. and Tapper, U. (1994) Positive matrix factorization: A non-negative factor model with optimal utilization of error. *Environmetrics* **5**(1): 111-126.

Perron, O. (1907) Zur Theorie der Matrices. *Mathematische Annalen* **64**(2): 248-263.

Roy, O. and Vetterli, M. (2007) The effective rank: A measure of effective dimensionality. In: *European Signal Processing Conference (EUSIPCO).* Poznań, Poland (September 3-7, 2007), pp. 606-610.

Shahnaz, F., Berry, M.W., Pauca, V.P. and Plemmons, R.J. (2006) Document clustering using nonnegative matrix factorization. *Information Processing and Management* **42**(2): 373-386.

Steinhaus, H. (1957) Sur la division des corps matériels en parties. *Bull. Acad. Polon. Sci.* **4**(12): 801-804.

Svensson, L.E.O. (1994) Estimating and interpreting forward interest rates: Sweden 1992-1994. *NBER Working Paper No. 4871*. Cambridge, MA: National Bureau of Economic Research.

Takada, H.H. and Stern, J.M. (2015) Non-negative matrix factorization and term structure of interest rates. *AIP Conference Proceedings* **1641**(1): 369-377.

Yang, W., Gibson, J.D. and He, T. (2005) Coefficient rate and lossy source coding. *IEEE Transactions on Information Theory* **51**(1): 381-386.

Zass, R. and Shashua, A. (2005) A unifying approach to hard and probabilistic clustering. In: *Proceedings of the Tenth IEEE International Conference on Computer Vision (ICCV'05).* Washington, DC: IEEE Computer Society, pp. 294-301.

**Table 1.** The weights matrix $W_{iA}$ (in %, rounded to 2 decimals) averaged over $P = 100$ vanilla NMF runs for $K = 2$. W1 and W2 correspond to the first and second columns of $W_{iA}$. SD1 and SD2 are the corresponding standard deviations. W1 and W2 are plotted against the log of the maturity in Figures 3 and 4. When clustering the columns of the matrix $\widehat{W}$ (whose dimension in this case is $12 \times 200$ – see Section 2), we get 2 batches with $P_1 = P_2 = 100$ elements in each.

| Maturity | W1 | W2 | SD1 | SD2 |
|---|---|---|---|---|
| 1 Mo | 12.2 | 4.27 | 2.16 | 1.73 |
| 2 Mo | 11.8 | 4.69 | 1.95 | 1.56 |
| 3 Mo | 11.32 | 5.12 | 1.7 | 1.36 |
| 6 Mo | 10.27 | 6.26 | 1.09 | 0.88 |
| 1 Yr | 8.24 | 7.95 | 0.08 | 0.06 |
| 2 Yr | 5.86 | 9.7 | 1.05 | 0.84 |
| 3 Yr | 5.15 | 10.17 | 1.37 | 1.1 |
| 5 Yr | 5.02 | 10.41 | 1.47 | 1.18 |
| 7 Yr | 5.55 | 10.54 | 1.36 | 1.09 |
| 10 Yr | 6.15 | 10.66 | 1.23 | 0.99 |
| 20 Yr | 8.39 | 10.27 | 0.51 | 0.41 |
| 30 Yr | 10.04 | 9.96 | 0.02 | 0.02 |

**Table 2.** The fit measures for the same NMF runs as in Table 1. See Subsection 2.2 for the definitions of $\rho_i$ and $E_i$. All values are rounded to 2 decimals. In this and other tables below, $\rho_i$ are expressed in %, while $E_i$ are the actual values of the errors (not in %).

| Maturity | Correlations $\rho_i$ | Errors $E_i$ |
|---|---|---|
| 1 Mo | 98.11 | 0.71 |
| 2 Mo | 98.8 | 0.51 |
| 3 Mo | 99.12 | 0.48 |
| 6 Mo | 99.68 | 0.45 |
| 1 Yr | 99.23 | 0.86 |
| 2 Yr | 99.55 | 0.56 |
| 3 Yr | 99.69 | 0.47 |
| 5 Yr | 99.81 | 0.35 |
| 7 Yr | 99.89 | 0.27 |
| 10 Yr | 99.81 | 0.4 |
| 20 Yr | 99.21 | 1.37 |
| 30 Yr | 98.3 | 2.38 |

**Table 3.** The weights matrix $W_{iA}$ (in %, rounded to 2 decimals) averaged over $P = 100$ vanilla NMF runs for $K = 3$. W1, W2 and W3 correspond to the first, second and third columns of $W_{iA}$. SD1, SD2 and SD3 are the corresponding standard deviations. W1, W2 and W3 are plotted against the log of the maturity in Figures 8, 9 and 10. When clustering the columns of the matrix $\widehat{W}$ (whose dimension in this case is $12 \times 300$ – see Section 2), we get 3 batches with nonuniform numbers of elements $P_A$ in each batch, which vary from one set of $P = 100$ runs to another. For the set reported in this table we have $P_1 = 91$, $P_2 = 97$ and $P_3 = 112$. As mentioned in Section 2, this is indicative of instability for $K = 3$ (owing to the "overall mode").

| Maturity | W1    | W2   | W3    | SD1  | SD2  | SD3  |
|----------|-------|------|-------|------|------|------|
| 1 Mo     | 13.63 | 8.47 | 3.32  | 1.71 | 1.39 | 1.56 |
| 2 Mo     | 13.01 | 8.45 | 3.89  | 1.73 | 1.54 | 1.48 |
| 3 Mo     | 12.43 | 8.44 | 4.37  | 1.49 | 1.17 | 1.31 |
| 6 Mo     | 11.04 | 8.49 | 5.61  | 1.72 | 1.37 | 1.33 |
| 1 Yr     | 8.5   | 8.26 | 7.59  | 1.88 | 1.81 | 1.74 |
| 2 Yr     | 5.23  | 7.93 | 9.89  | 1.67 | 1.81 | 1.58 |
| 3 Yr     | 4.4   | 7.64 | 10.58 | 1.44 | 1.66 | 1.52 |
| 5 Yr     | 4.31  | 7.49 | 10.99 | 1.92 | 1.69 | 1.41 |
| 7 Yr     | 4.66  | 7.89 | 11.15 | 1.27 | 1.29 | 1.08 |
| 10 Yr    | 5.38  | 8.03 | 11.37 | 1.47 | 1.37 | 1.4  |
| 20 Yr    | 7.77  | 9.09 | 10.84 | 2.42 | 2.53 | 2.43 |
| 30 Yr    | 9.64  | 9.83 | 10.39 | 3.19 | 3.15 | 2.88 |

**Table 4.** The fit measures for the same NMF runs as in Table 3. See Subsection 2.2 for the definitions of $\rho_i$ and $E_i$. All values are rounded to 2 decimals.

| Maturity | Correlations $\rho_i$ | Errors $E_i$ |
|----------|----------------------|--------------|
| 1 Mo     | 99.27                | 2.7          |
| 2 Mo     | 99.62                | 2.03         |
| 3 Mo     | 99.49                | 1.81         |
| 6 Mo     | 99.42                | 1.24         |
| 1 Yr     | 99.24                | 1.7          |
| 2 Yr     | 99.62                | 1.93         |
| 3 Yr     | 99.76                | 2.17         |
| 5 Yr     | 99.92                | 1.92         |
| 7 Yr     | 99.91                | 1.88         |
| 10 Yr    | 99.79                | 1.83         |
| 20 Yr    | 98.65                | 1.86         |
| 30 Yr    | 97.77                | 2.45         |

**Table 5.** The weights matrix $W_{iA}$ (in %, rounded to 2 decimals) averaged over $P = 100$ de-noised NMF runs for $K = 2$. W1 and W2 correspond to the first and second columns of $W_{iA}$. SD1 and SD2 (also in % but rounded to 6 decimals) are the corresponding standard deviations. W1 and W2 are plotted against the log of the maturity in Figures 13 and 14. When clustering the columns of the matrix $\widehat{W}$ (whose dimension in this case is $12 \times 200$ – see Section 2), we get 2 batches with $P_1 = P_2 = 100$ elements in each.

| Maturity | W1 | W2 | SD1 | SD2 |
|---|---|---|---|---|
| 1 Mo | 0 | 12.17 | 0 | 0.000145 |
| 2 Mo | 0.8 | 11.82 | 5.1e-05 | 0.000135 |
| 3 Mo | 1.44 | 11.09 | 3.3e-05 | 0.000114 |
| 6 Mo | 3.95 | 10.12 | 2e-05 | 7.1e-05 |
| 1 Yr | 6.32 | 6.29 | 1.8e-05 | 2e-05 |
| 2 Yr | 7.38 | 1.43 | 7.4e-05 | 0.000146 |
| 3 Yr | 7.16 | 0.16 | 0.000138 | 0.000228 |
| 5 Yr | 8.26 | 0 | 9.1e-05 | 0 |
| 7 Yr | 10.83 | 2.86 | 2.8e-05 | 7.6e-05 |
| 10 Yr | 13.42 | 6.11 | 4.1e-05 | 4.6e-05 |
| 20 Yr | 18.47 | 15.54 | 3.5e-05 | 7e-06 |
| 30 Yr | 21.98 | 22.41 | 4.7e-05 | 5.5e-05 |

**Table 6.** The fit measures for the same NMF runs as in Table 5 for the de-noised matrix $Z_{is}$. See Subsection 2.2 for the definitions of $\rho_i$ and $E_i$. All values are rounded to 2 decimals.

| Maturity | Correlations $\rho_i$ | Errors $E_i$ |
|---|---|---|
| 1 Mo | 96.48 | 0.92 |
| 2 Mo | 97.15 | 0.67 |
| 3 Mo | 96.51 | 0.75 |
| 6 Mo | 95.22 | 0.5 |
| 1 Yr | 91.31 | 0.83 |
| 2 Yr | 97.53 | 1.07 |
| 3 Yr | 97.38 | 1.92 |
| 5 Yr | 97.43 | 1.47 |
| 7 Yr | 98.95 | 0.47 |
| 10 Yr | 98.17 | 0.58 |
| 20 Yr | 92.97 | 4.85 |
| 30 Yr | 88.95 | 11.75 |

**Table 7.** The weights matrix $W_{iA}$ (in %, rounded to 2 decimals) averaged over $P = 100$ de-noised NMF runs for $K = 3$. W1, W2 and W3 correspond to the first, second and third columns of $W_{iA}$. SD1, SD2 and SD3 (also in % but rounded to 6 decimals) are the corresponding standard deviations. W1, W2 and W3 are plotted against the log of the maturity in Figures 18, 19 and 20. When clustering the columns of the matrix $\widehat{W}$ (whose dimension in this case is $12 \times 300$ – see Section 2), we get 3 batches with $P_1 = P_2 = P_3 = 100$ elements in each.

| Maturity | W1 | W2 | W3 | SD1 | SD2 | SD3 |
|---|---|---|---|---|---|---|
| 1 Mo | 17.17 | 0 | 0 | 0.518361 | 0 | 0 |
| 2 Mo | 16.78 | 1.28 | 0.02 | 0.507747 | 0.223593 | 0.005382 |
| 3 Mo | 15.83 | 2.34 | 0 | 0.477415 | 0.41493 | 0 |
| 6 Mo | 13.13 | 4.66 | 2.94 | 0.3091 | 0.306576 | 0.00226 |
| 1 Yr | 7.91 | 8.42 | 3.11 | 0.146807 | 0.942744 | 0.00286 |
| 2 Yr | 1.84 | 11.73 | 1.23 | 0.020332 | 1.865025 | 0.003985 |
| 3 Yr | 0 | 12.92 | 0 | 0 | 2.29633 | 0 |
| 5 Yr | 0 | 12.59 | 1.61 | 0 | 1.932447 | 0.05651 |
| 7 Yr | 1.28 | 12.65 | 7.79 | 0.192364 | 0.861373 | 0.004615 |
| 10 Yr | 2.7 | 12.3 | 14.7 | 0.402724 | 0.424659 | 0.039581 |
| 20 Yr | 9.37 | 11.1 | 29.01 | 0.599507 | 3.177687 | 0.012721 |
| 30 Yr | 13.99 | 10.01 | 39.59 | 0.785305 | 5.240601 | 0.025809 |

**Table 8.** The fit measures for the same NMF runs as in Table 7 for the de-noised matrix $Z_{is}$. See Subsection 2.2 for the definitions of $\rho_i$ and $E_i$. All values are rounded to 2 decimals.

| Maturity | Correlations $\rho_i$ | Errors $E_i$ |
|---|---|---|
| 1 Mo | 99.35 | 0.15 |
| 2 Mo | 99.35 | 0.13 |
| 3 Mo | 99.06 | 0.16 |
| 6 Mo | 97.47 | 0.23 |
| 1 Yr | 93.99 | 0.58 |
| 2 Yr | 99.64 | 0.11 |
| 3 Yr | 99.87 | 0.07 |
| 5 Yr | 99.6 | 0.13 |
| 7 Yr | 99.6 | 0.13 |
| 10 Yr | 99.65 | 0.12 |
| 20 Yr | 99.24 | 0.18 |
| 30 Yr | 97.41 | 0.43 |

**Table 9.** Correlations $\varphi_{AB}$ and $\vartheta_A$ (in %) for $K = 3$ across 5 different sets of $P = 100$ de-noised NMF runs (see Subsection 2.3 for details). The factors are in the same order as in Table 7 for sets 1-3 as for these sets the ordering is evident from the $\vartheta_1$, $\vartheta_2$ and $\varphi_{12}$ correlations (and the first set is the same as in Table 7). However, for sets 4 and 5 the factors are not ordered in any way as no ordering is evident from the correlations.

| Set | $\vartheta_1$ | $\vartheta_2$ | $\vartheta_3$ | $\varphi_{12}$ | $\varphi_{13}$ | $\varphi_{23}$ |
|---|---|---|---|---|---|---|
| 1 | -69.92 | 34.25 | 48.28 | -48.23 | -72.25 | 18.18 |
| 2 | -69.92 | 34.25 | 41.63 | -48.23 | -62.97 | -3.22 |
| 3 | -69.91 | 34.26 | 45.72 | -48.24 | -68.73 | 9.83 |
| 4 | 22.75 | 55.38 | -10.95 | 42.99 | -75.22 | 12.18 |
| 5 | -11.73 | 55.39 | 24.75 | 17.12 | -73.07 | 40.88 |

**Table 10.** The weights matrix $W_{iA}$ (in %, rounded to 2 decimals) averaged over $P = 100$ NMF runs for $K = 2$ with alternative de-noising (see Subsection 2.5). W1 and W2 correspond to the first and second columns of $W_{iA}$. SD1 and SD2 (also in % but rounded to 6 decimals) are the corresponding standard deviations. W1 and W2 are plotted against the log of the maturity in Figures 23 and 24. When clustering the columns of the matrix $\widehat{W}$ (whose dimension in this case is $12 \times 200$ – see Section 2), we get 2 batches with $P_1 = P_2 = 100$ elements in each.

| Maturity | W1 | W2 | SD1 | SD2 |
|---|---|---|---|---|
| 1 Mo | 18.12 | 0 | 0.014807 | 0 |
| 2 Mo | 16.8 | 1.24 | 0.012012 | 0.002196 |
| 3 Mo | 16.01 | 2.2 | 0.010738 | 0.001476 |
| 6 Mo | 13.25 | 4.36 | 0.006912 | 0.00078 |
| 1 Yr | 9.87 | 9.16 | 0.000512 | 9.9e-05 |
| 2 Yr | 6.79 | 15.13 | 0.006556 | 0.000659 |
| 3 Yr | 5.96 | 17.05 | 0.008713 | 0.000885 |
| 5 Yr | 5.29 | 17.2 | 0.009404 | 0.000932 |
| 7 Yr | 4.03 | 15.16 | 0.008822 | 0.000858 |
| 10 Yr | 2.82 | 12.8 | 0.007872 | 0.000784 |
| 20 Yr | 1.07 | 5.56 | 0.003543 | 0.000352 |
| 30 Yr | 0 | 0.14 | 0 | 0.000135 |

**Table 11.** The fit measures for the same NMF runs as in Table 10 for the matrix $\widetilde{Z}_{is}$ with alternative de-noising (see Subsection 2.5). See Subsection 2.2 for the definitions of $\rho_i$ and $E_i$. All values are rounded to 2 decimals.

| Maturity | Correlations $\rho_i$ | Errors $E_i$ |
|---|---|---|
| 1 Mo | 99.63 | 0.18 |
| 2 Mo | 99.76 | 0.1 |
| 3 Mo | 99.57 | 0.14 |
| 6 Mo | 98.94 | 0.2 |
| 1 Yr | 94.91 | 0.55 |
| 2 Yr | 97.25 | 0.2 |
| 3 Yr | 98.18 | 0.14 |
| 5 Yr | 99.32 | 0.06 |
| 7 Yr | 99.23 | 0.06 |
| 10 Yr | 97.6 | 0.14 |
| 20 Yr | 92.7 | 0.09 |
| 30 Yr | 28.63 | 0.09 |

**Table 12.** Sample data from the `treasuty.txt` file (see Appendix A).

| Date | 1 Mo | 2 Mo | 3 Mo | 6 Mo | 1 Yr | 2 Yr | 3 Yr | 5 Yr | 7 Yr | 10 Yr | 20 Yr | 30 Yr |
|---|---|---|---|---|---|---|---|---|---|---|---|---|
| 10/25/19 | 1.73 | 1.72 | 1.66 | 1.66 | 1.60 | 1.63 | 1.62 | 1.62 | 1.71 | 1.80 | 2.10 | 2.29 |
| 10/28/19 | 1.74 | 1.71 | 1.65 | 1.65 | 1.60 | 1.64 | 1.65 | 1.66 | 1.75 | 1.85 | 2.16 | 2.34 |
| 10/29/19 | 1.66 | 1.67 | 1.63 | 1.64 | 1.59 | 1.64 | 1.65 | 1.66 | 1.74 | 1.84 | 2.15 | 2.33 |
| 10/30/19 | 1.61 | 1.60 | 1.62 | 1.62 | 1.59 | 1.61 | 1.60 | 1.61 | 1.69 | 1.78 | 2.08 | 2.26 |
| 10/31/19 | 1.59 | 1.59 | 1.54 | 1.57 | 1.53 | 1.52 | 1.52 | 1.51 | 1.60 | 1.69 | 2.00 | 2.17 |
| 11/01/19 | 1.58 | 1.58 | 1.52 | 1.55 | 1.53 | 1.56 | 1.55 | 1.55 | 1.63 | 1.73 | 2.03 | 2.21 |
| 11/04/19 | 1.58 | 1.57 | 1.53 | 1.57 | 1.56 | 1.60 | 1.59 | 1.60 | 1.69 | 1.79 | 2.10 | 2.27 |
| 11/05/19 | 1.56 | 1.57 | 1.56 | 1.58 | 1.62 | 1.63 | 1.63 | 1.66 | 1.77 | 1.86 | 2.17 | 2.34 |
| 11/06/19 | 1.55 | 1.56 | 1.56 | 1.57 | 1.58 | 1.61 | 1.60 | 1.63 | 1.73 | 1.81 | 2.13 | 2.30 |
| 11/07/19 | 1.57 | 1.57 | 1.56 | 1.58 | 1.58 | 1.68 | 1.70 | 1.74 | 1.84 | 1.92 | 2.24 | 2.40 |
| 11/08/19 | 1.56 | 1.56 | 1.55 | 1.58 | 1.58 | 1.68 | 1.70 | 1.74 | 1.86 | 1.94 | 2.27 | 2.43 |
| 11/12/19 | 1.56 | 1.56 | 1.59 | 1.59 | 1.58 | 1.66 | 1.69 | 1.73 | 1.84 | 1.92 | 2.24 | 2.39 |
| 11/13/19 | 1.56 | 1.57 | 1.57 | 1.59 | 1.57 | 1.63 | 1.65 | 1.69 | 1.79 | 1.88 | 2.20 | 2.36 |
| 11/14/19 | 1.59 | 1.57 | 1.57 | 1.58 | 1.55 | 1.58 | 1.59 | 1.63 | 1.73 | 1.82 | 2.15 | 2.31 |
| 11/15/19 | 1.59 | 1.56 | 1.57 | 1.59 | 1.54 | 1.61 | 1.61 | 1.65 | 1.75 | 1.84 | 2.16 | 2.31 |
| 11/18/19 | 1.59 | 1.57 | 1.57 | 1.58 | 1.54 | 1.60 | 1.59 | 1.63 | 1.73 | 1.81 | 2.14 | 2.30 |
| 11/19/19 | 1.58 | 1.57 | 1.57 | 1.58 | 1.54 | 1.60 | 1.59 | 1.63 | 1.71 | 1.79 | 2.11 | 2.26 |
| 11/20/19 | 1.57 | 1.56 | 1.57 | 1.58 | 1.54 | 1.56 | 1.55 | 1.58 | 1.66 | 1.73 | 2.05 | 2.20 |
| 11/21/19 | 1.57 | 1.57 | 1.58 | 1.59 | 1.55 | 1.60 | 1.59 | 1.62 | 1.71 | 1.77 | 2.09 | 2.24 |
| 11/22/19 | 1.58 | 1.59 | 1.58 | 1.59 | 1.56 | 1.61 | 1.60 | 1.62 | 1.71 | 1.77 | 2.08 | 2.22 |

**Table 13.** The weights matrix $W_{iA}$ (in %, rounded to 2 decimals) using the statistical clustering approach of Section 3 for $K = 2$ clusters. The clustering was determined using 100 sets of 100 k-means runs (see Section 3) and was 100% stable from set to set. W1 and W2 correspond to the first and second columns of $W_{iA}$.

| Maturity | W1 | W2 |
| --- | --- | --- |
| 1 Mo | 24.82 | 0 |
| 2 Mo | 24.93 | 0 |
| 3 Mo | 24.94 | 0 |
| 6 Mo | 25.3 | 0 |
| 1 Yr | 0 | 11.9 |
| 2 Yr | 0 | 11.67 |
| 3 Yr | 0 | 11.57 |
| 5 Yr | 0 | 11.67 |
| 7 Yr | 0 | 12.13 |
| 10 Yr | 0 | 12.63 |
| 20 Yr | 0 | 13.79 |
| 30 Yr | 0 | 14.64 |

**Table 14.** The fit measures for the clustering runs in Table 13. See Subsection 2.2 for the definitions of $\rho_i$ and $E_i$. All values are rounded to 2 decimals.

| Maturity | Correlations $\rho_i$ | Errors $E_i$ |
| --- | --- | --- |
| 1 Mo | 98.26 | 0.82 |
| 2 Mo | 99.55 | 0.22 |
| 3 Mo | 99.79 | 0.1 |
| 6 Mo | 98.04 | 1.48 |
| 1 Yr | 96.91 | 3.32 |
| 2 Yr | 99.61 | 0.88 |
| 3 Yr | 99.64 | 1.66 |
| 5 Yr | 99.79 | 1.78 |
| 7 Yr | 99.85 | 1.06 |
| 10 Yr | 99.87 | 0.58 |
| 20 Yr | 99.57 | 1.5 |
| 30 Yr | 99.18 | 4.99 |

**Table 15.** The weights matrix $W_{iA}$ (in %, rounded to 2 decimals) using the statistical clustering approach of Section 3 for $K = 3$ clusters. The clustering was determined using 100 sets of 100 k-means runs (see Section 3) and was 100% stable from set to set. W1, W2 and W3 correspond to the first, second and third columns of $W_{iA}$.

| Maturity | W1 | W2 | W3 |
|---|---|---|---|
| 1 Mo | 33.23 | 0 | 0 |
| 2 Mo | 33.38 | 0 | 0 |
| 3 Mo | 33.39 | 0 | 0 |
| 6 Mo | 0 | 31.03 | 0 |
| 1 Yr | 0 | 30.93 | 0 |
| 2 Yr | 0 | 0 | 15.89 |
| 3 Yr | 0 | 0 | 15.76 |
| 5 Yr | 0 | 0 | 15.89 |
| 7 Yr | 0 | 0 | 16.51 |
| 10 Yr | 0 | 0 | 17.19 |
| 20 Yr | 0 | 0 | 18.76 |
| 30 Yr | 0 | 38.04 | 0 |

**Table 16.** The fit measures for the clustering runs in Table 15. See Subsection 2.2 for the definitions of $\rho_i$ and $E_i$. All values are rounded to 2 decimals.

| Maturity | Correlations $\rho_i$ | Errors $E_i$ |
|---|---|---|
| 1 Mo | 99.33 | 0.29 |
| 2 Mo | 99.9 | 0.04 |
| 3 Mo | 99.42 | 0.3 |
| 6 Mo | 96.77 | 2.12 |
| 1 Yr | 99.33 | 1.31 |
| 2 Yr | 99.59 | 0.5 |
| 3 Yr | 99.79 | 0.63 |
| 5 Yr | 99.94 | 0.6 |
| 7 Yr | 99.97 | 0.2 |
| 10 Yr | 99.89 | 0.15 |
| 20 Yr | 99.53 | 3.18 |
| 30 Yr | 97.74 | 2.22 |

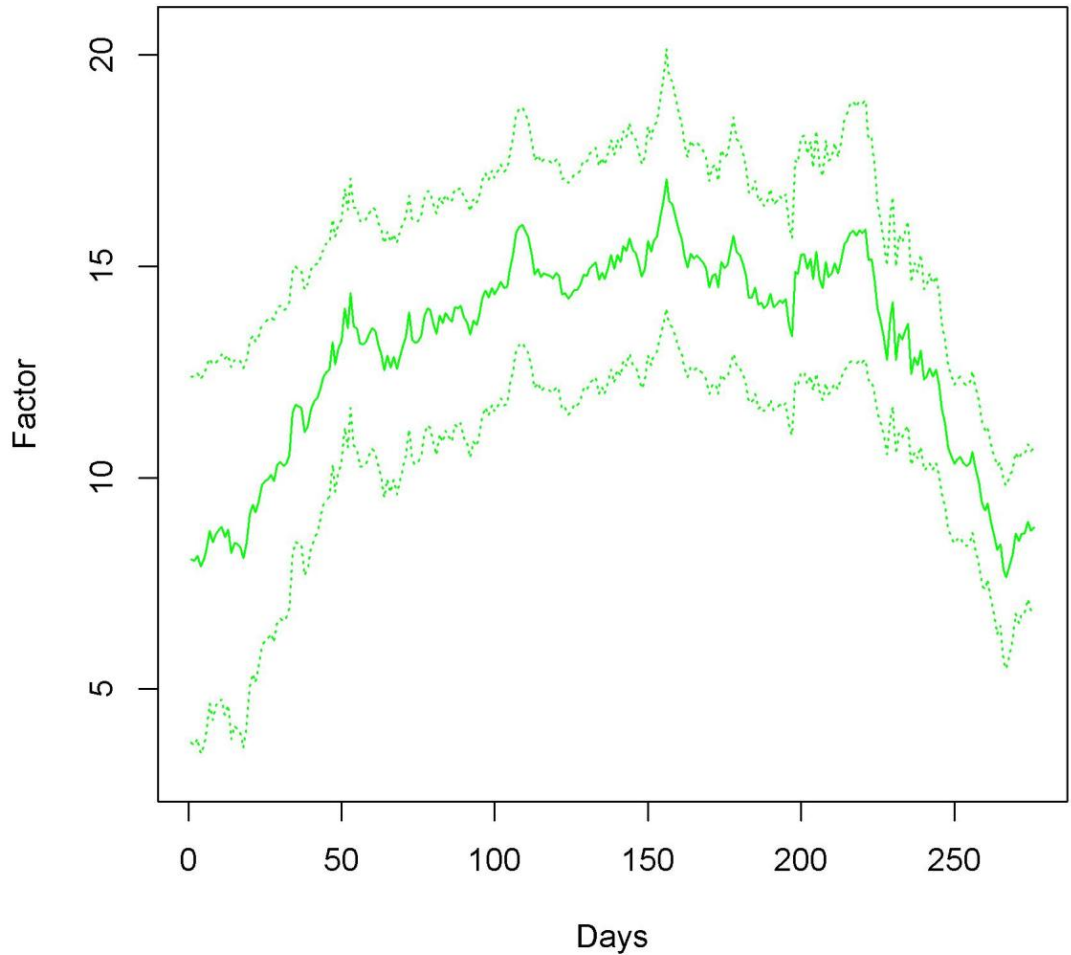

**Figure 1.** The time series plot for the first row of the factors matrix $F_{As}$ corresponding to the weights W1 in Table 1. The solid line is the mean over $P = 100$ runs, while the dashed lines correspond to one standard deviation from the mean in each direction.

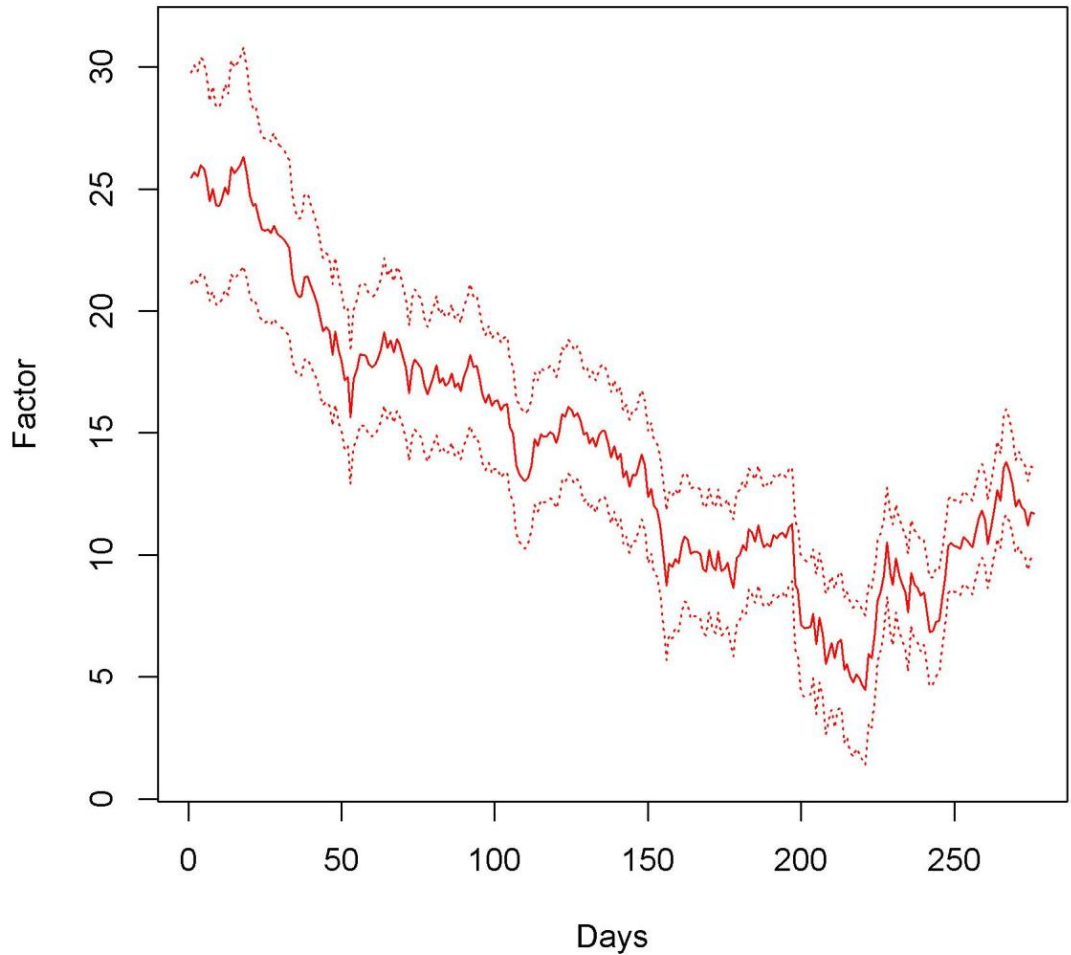

**Figure 2.** The time series plot for the second row of the factors matrix $F_{As}$ corresponding to the weights W2 in Table 1. The solid line is the mean over $P = 100$ runs, while the dashed lines correspond to one standard deviation from the mean in each direction.

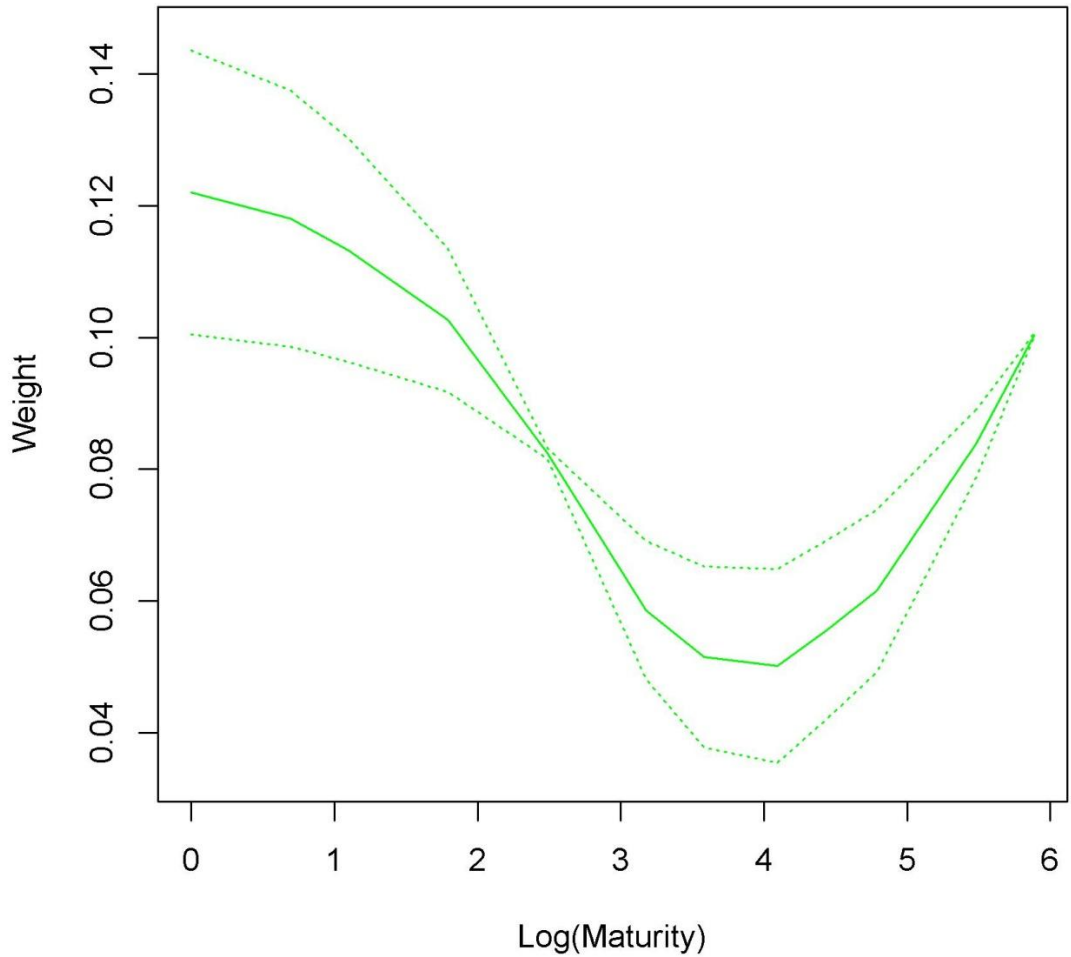

**Figure 3.** The weights W1 in Table 1 plotted against the natural log of the maturity. The solid line is the mean over $P = 100$ runs, while the dashed lines correspond to one standard deviation from the mean in each direction.

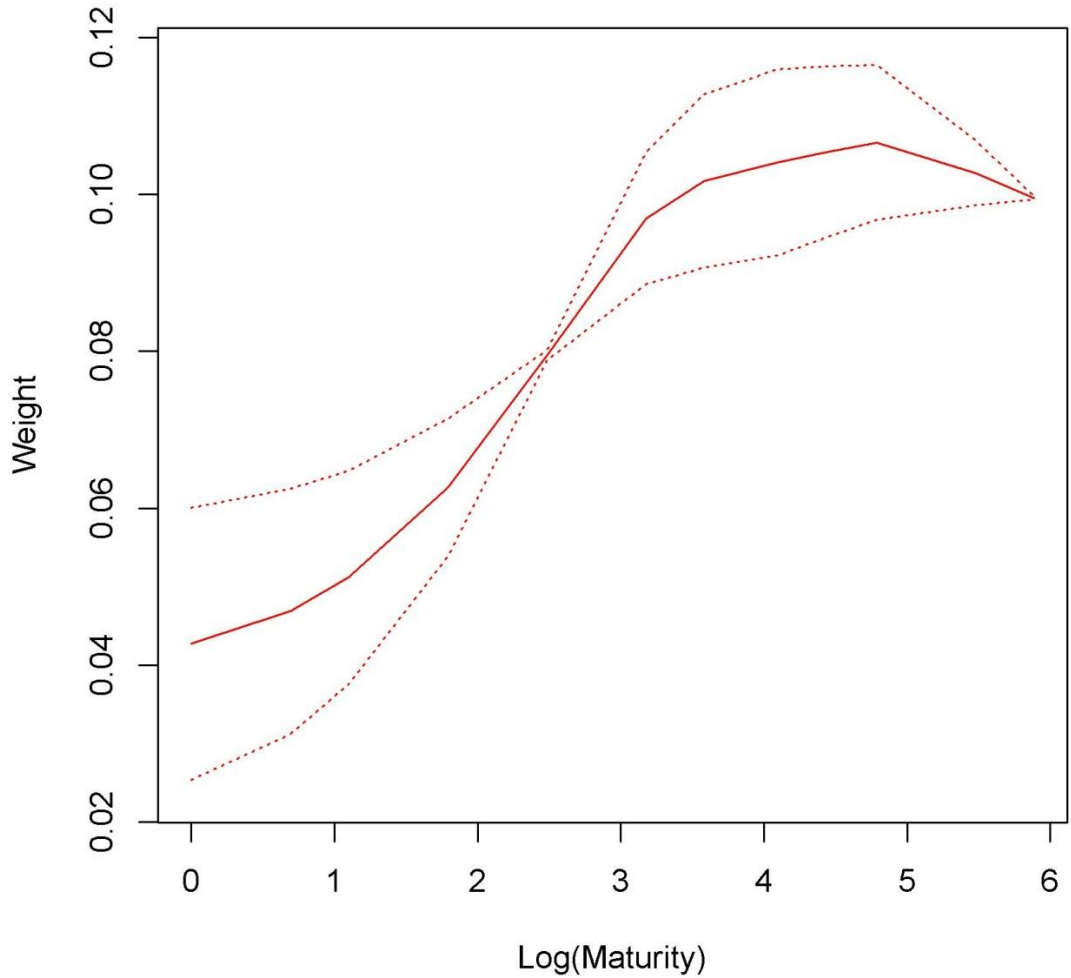

**Figure 4.** The weights W2 in Table 1 plotted against the natural log of the maturity. The solid line is the mean over $P = 100$ runs, while the dashed lines correspond to one standard deviation from the mean in each direction.

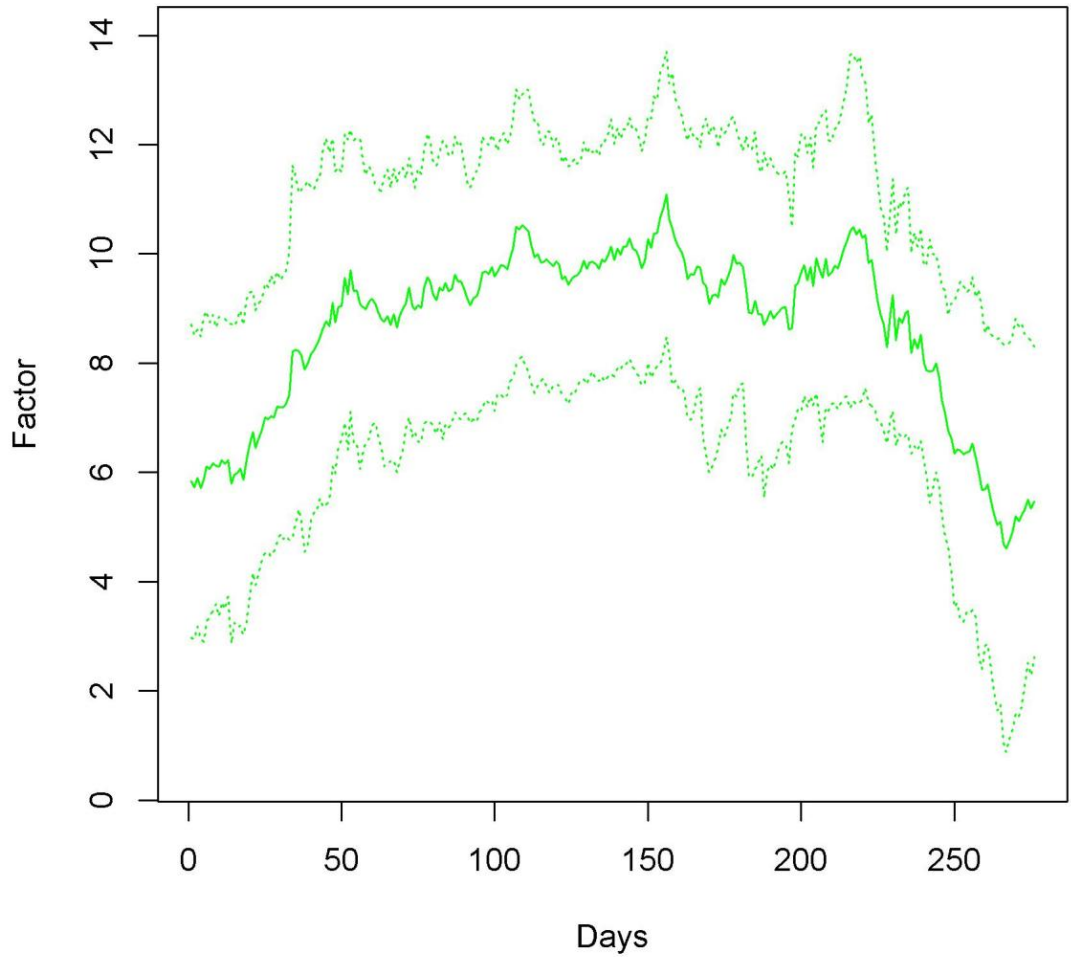

**Figure 5.** The time series plot for the first row of the factors matrix $F_{As}$ corresponding to the weights W1 in Table 3. The solid line is the mean over $P = 100$ runs, while the dashed lines correspond to one standard deviation from the mean in each direction.

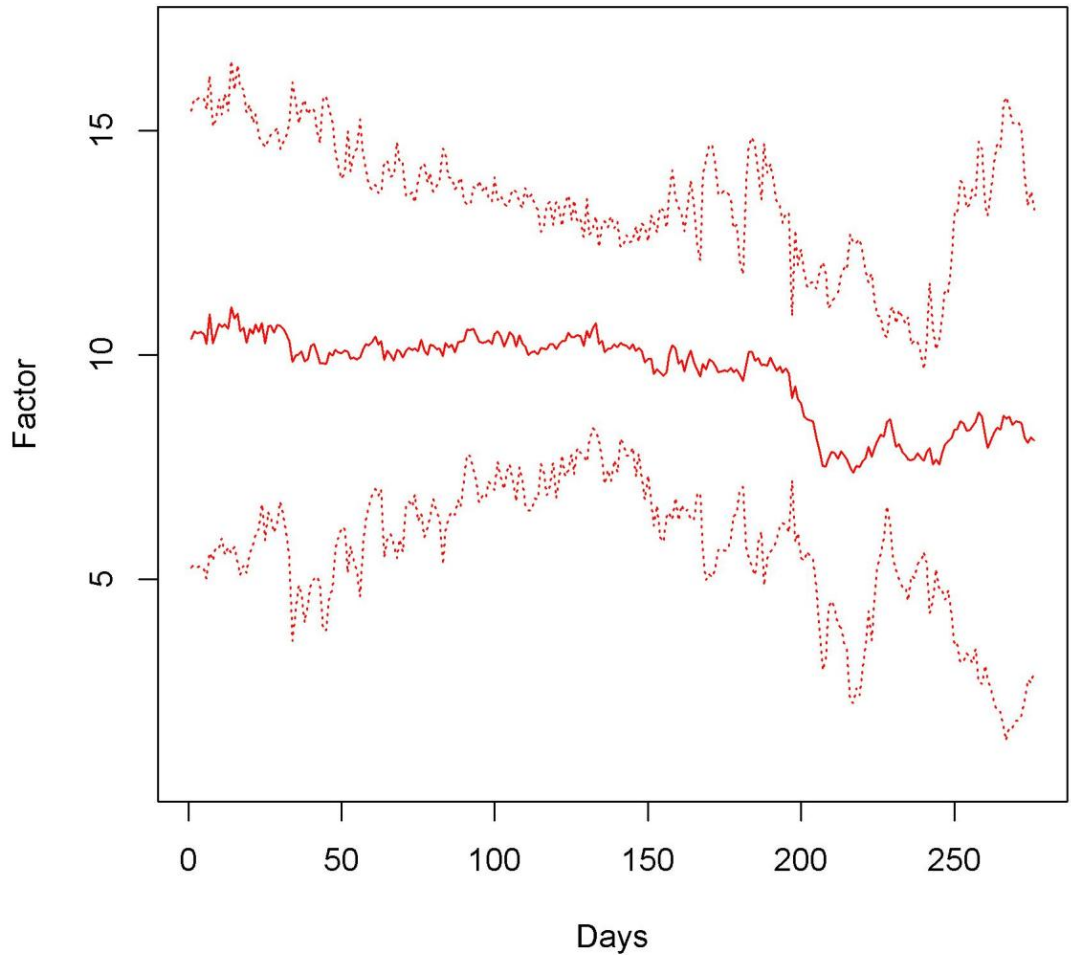

**Figure 6.** The time series plot for the second row of the factors matrix $F_{As}$ corresponding to the weights W2 in Table 3. The solid line is the mean over $P = 100$ runs, while the dashed lines correspond to one standard deviation from the mean in each direction.

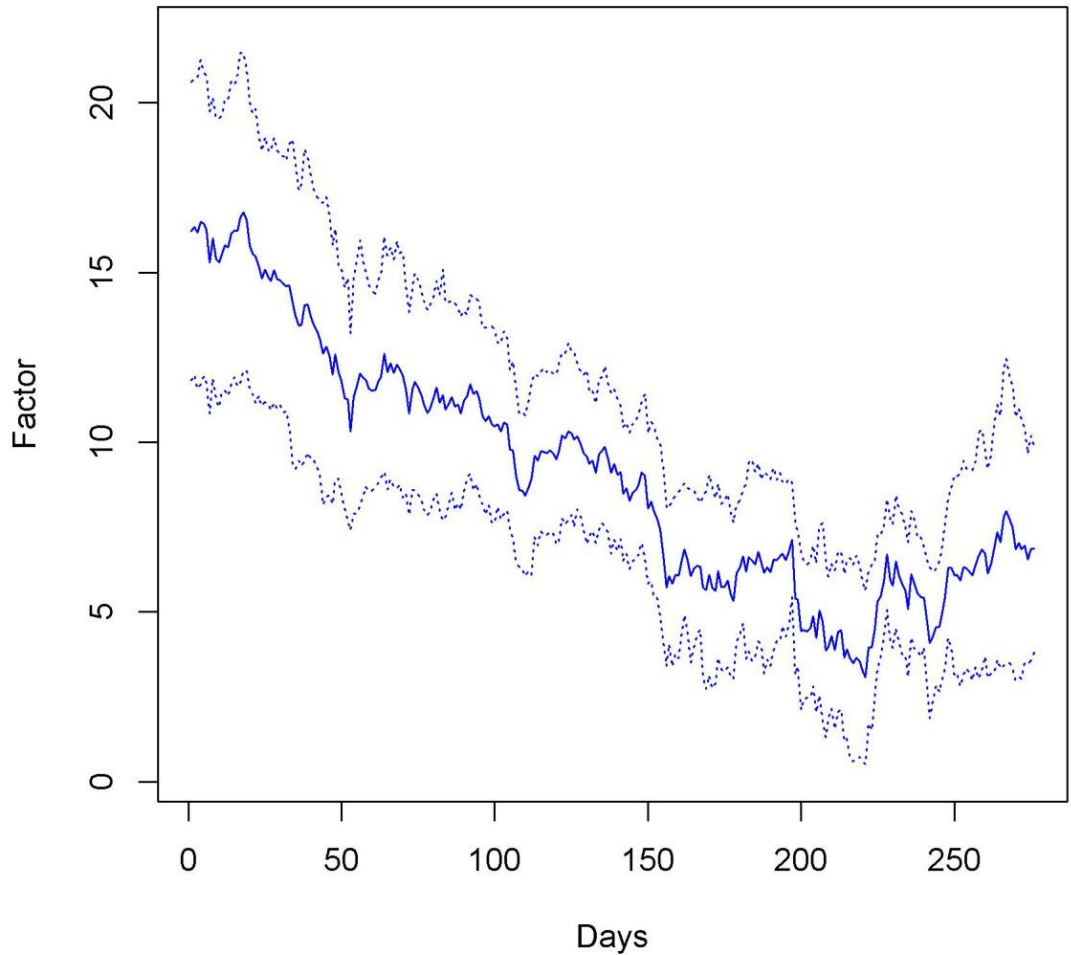

**Figure 7.** The time series plot for the third row of the factors matrix $F_{As}$ corresponding to the weights W3 in Table 3. The solid line is the mean over $P = 100$ runs, while the dashed lines correspond to one standard deviation from the mean in each direction.

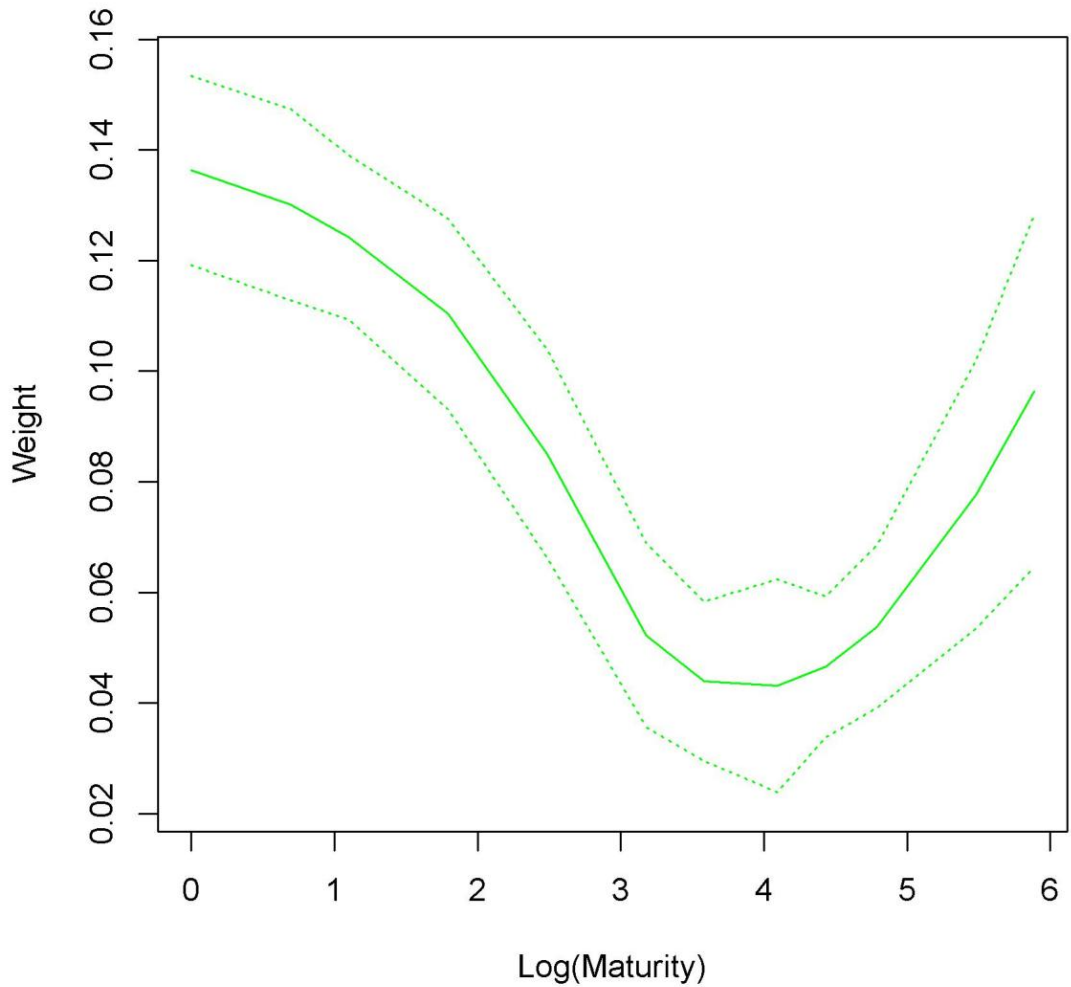

**Figure 8.** The weights W1 in Table 3 plotted against the natural log of the maturity. The solid line is the mean over $P = 100$ runs, while the dashed lines correspond to one standard deviation from the mean in each direction.

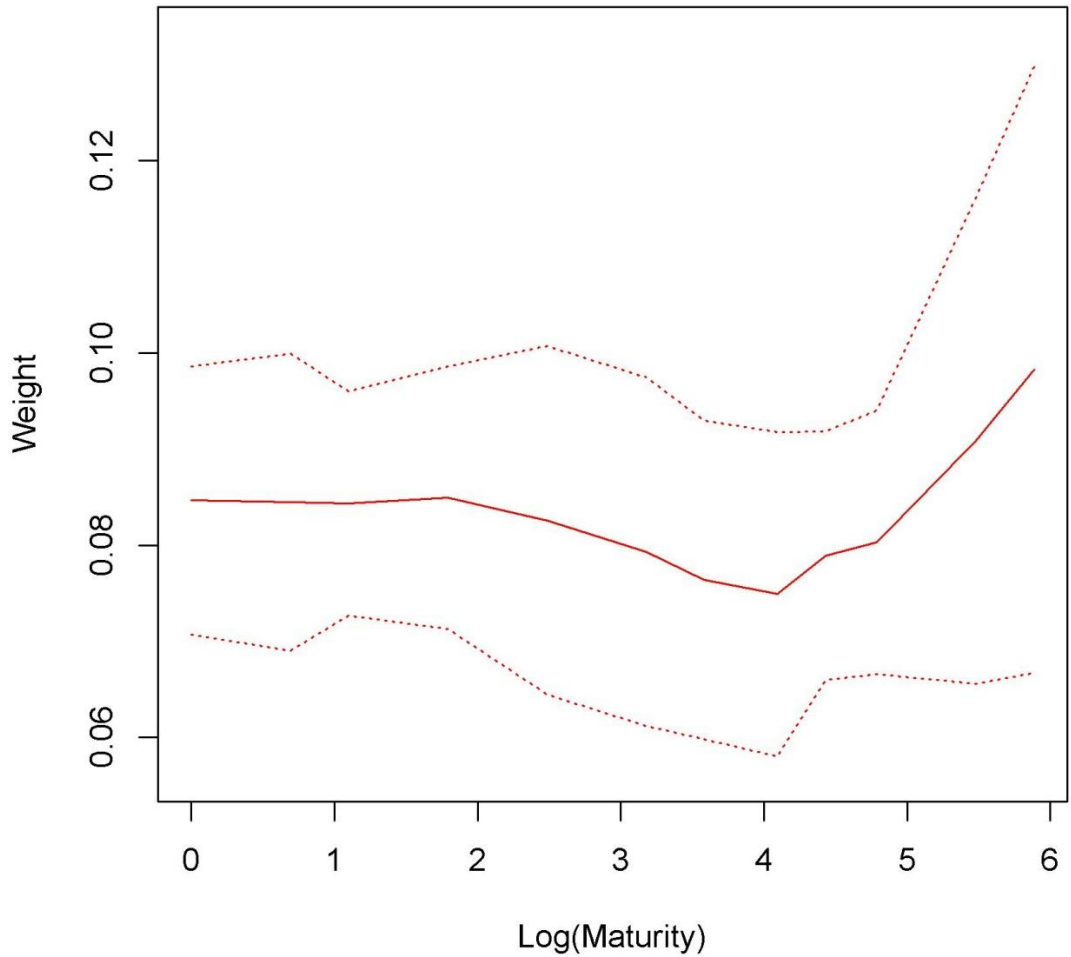

**Figure 9.** The weights W2 in Table 3 plotted against the natural log of the maturity. The solid line is the mean over $P = 100$ runs, while the dashed lines correspond to one standard deviation from the mean in each direction.

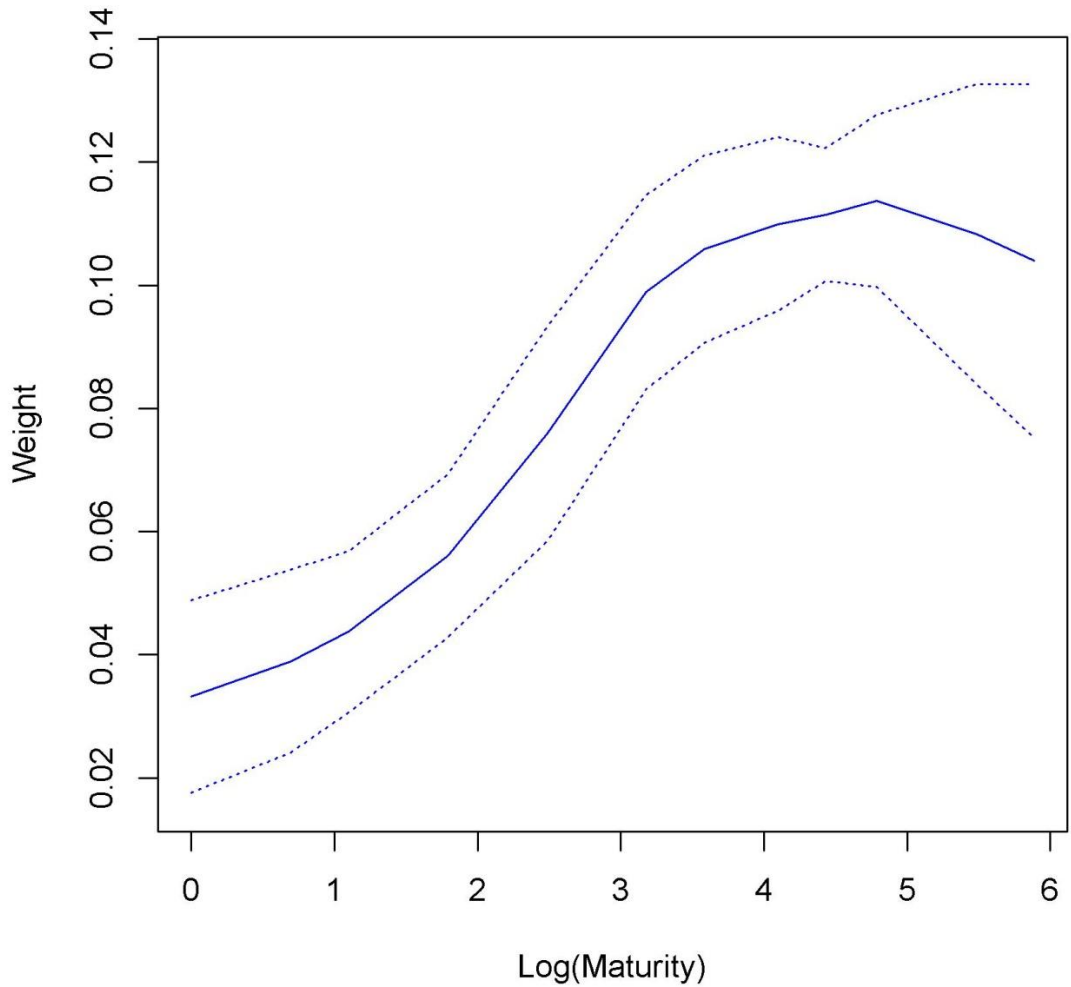

**Figure 10.** The weights W3 in Table 3 plotted against the natural log of the maturity. The solid line is the mean over $P = 100$ runs, while the dashed lines correspond to one standard deviation from the mean in each direction.

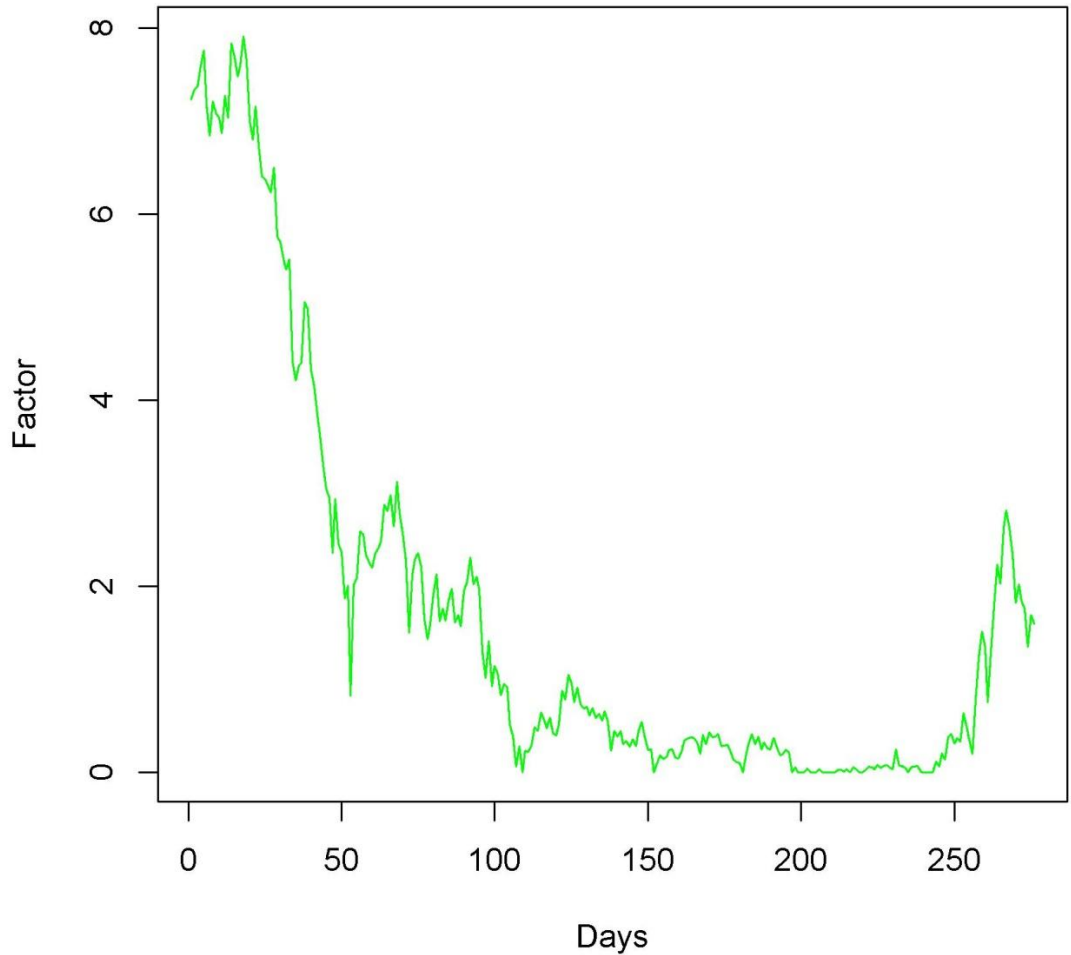

**Figure 11.** The time series plot for the first row of the factors matrix $F_{As}$ corresponding to the weights W1 in Table 5. The solid line is the mean over $P = 100$ runs. The dashed lines corresponding to one standard deviation from the mean in each direction are also included but are not visible in this graph as the errors are tiny (see Table 5).

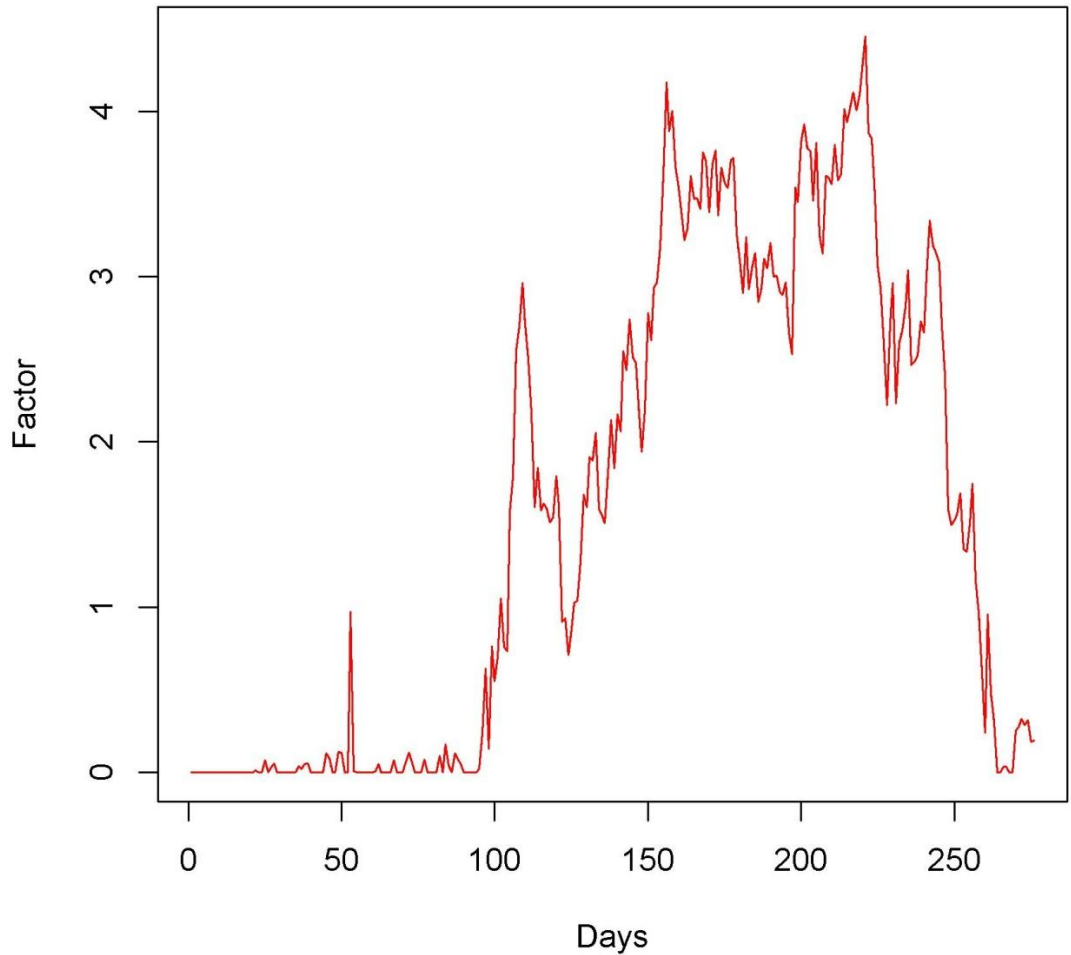

**Figure 12.** The time series plot for the second row of the factors matrix $F_{As}$ corresponding to the weights W2 in Table 5. The solid line is the mean over $P = 100$ runs. The dashed lines corresponding to one standard deviation from the mean in each direction are also included but are not visible in this graph as the errors are tiny (see Table 5).

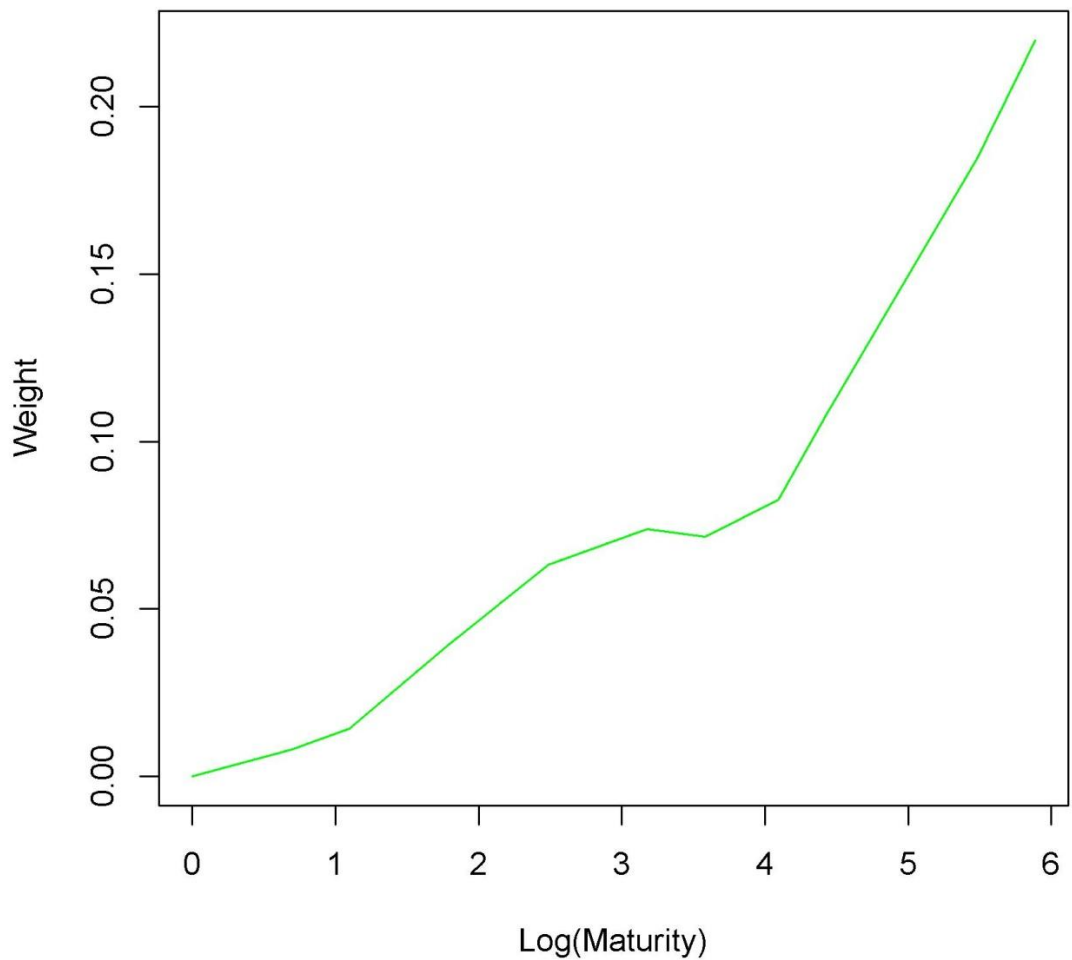

**Figure 13.** The weights W1 in Table 5 plotted against the natural log of the maturity. The solid line is the mean over $P = 100$ runs. The dashed lines corresponding to one standard deviation from the mean in each direction are also included but are not visible in this graph as the errors are tiny (see Table 5).

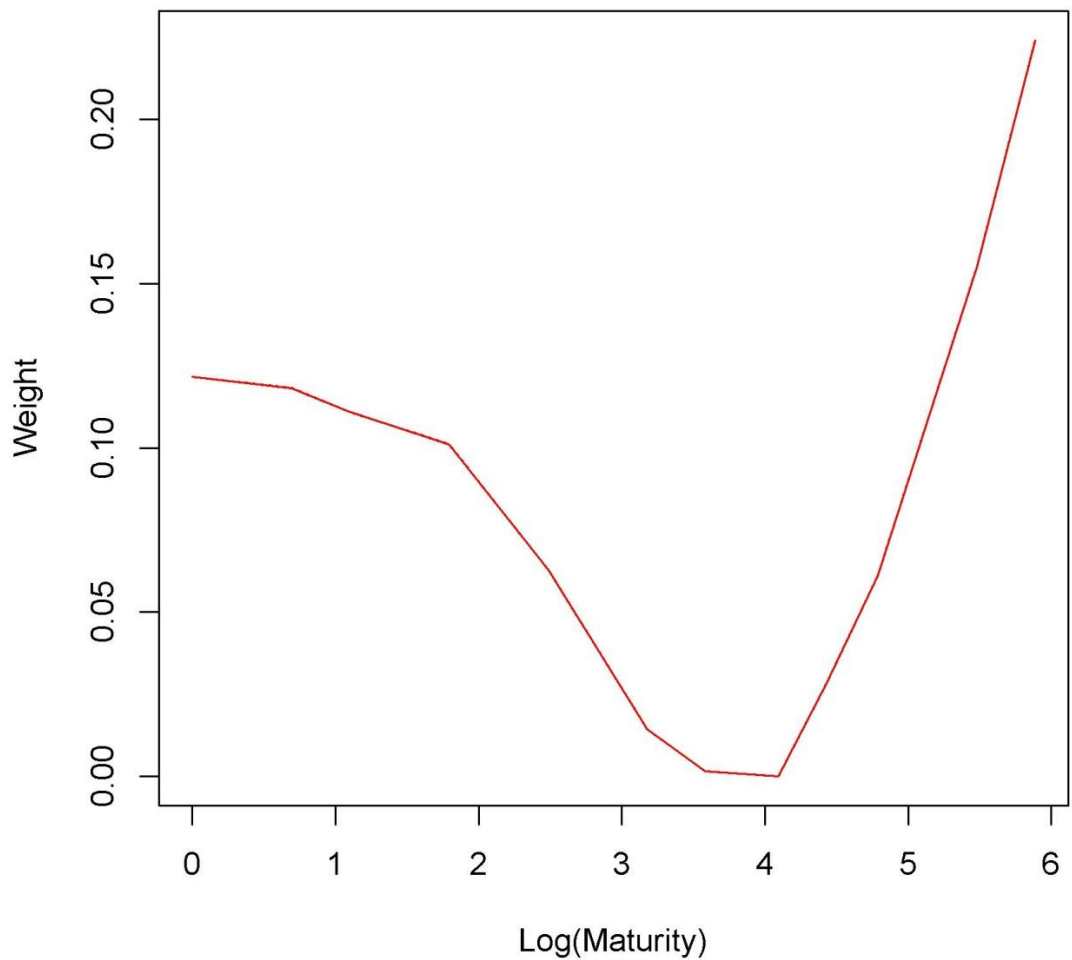

**Figure 14.** The weights W2 in Table 5 plotted against the natural log of the maturity. The solid line is the mean over $P = 100$ runs. The dashed lines corresponding to one standard deviation from the mean in each direction are also included but are not visible in this graph as the errors are tiny (see Table 5).

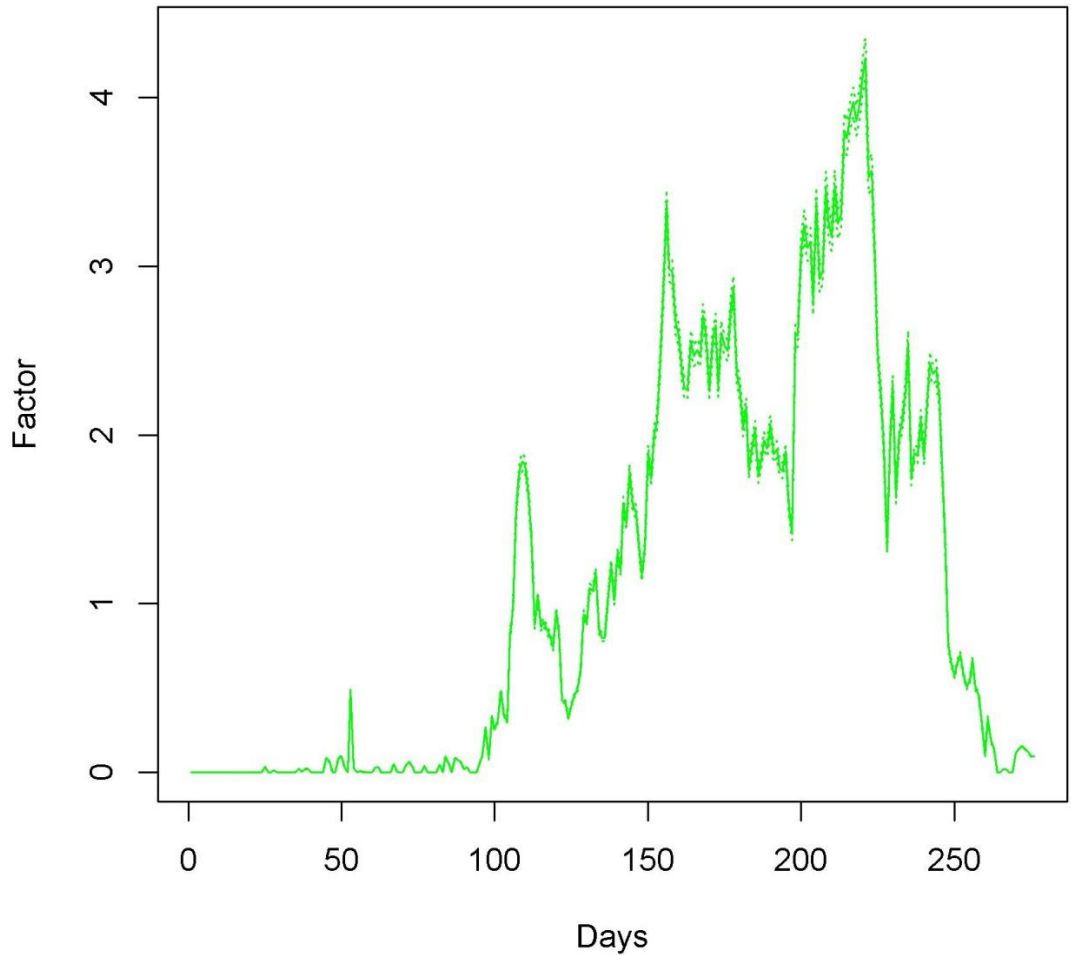

**Figure 15.** The time series plot for the first row of the factors matrix $F_{As}$ corresponding to the weights W1 in Table 7. The solid line is the mean over $P = 100$ runs. The dashed lines corresponding to one standard deviation from the mean in each direction are also included but are mostly not visible in this graph as the errors are relatively small.

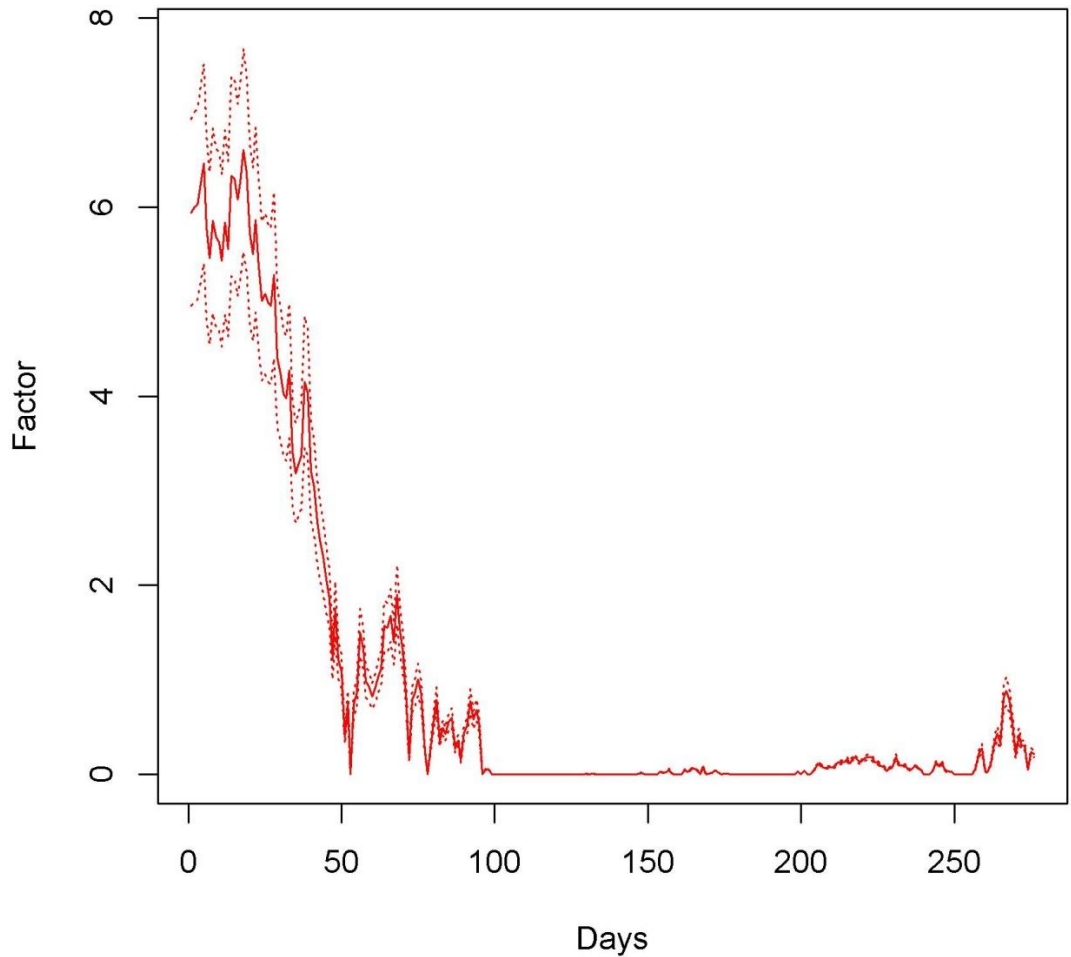

**Figure 16.** The time series plot for the second row of the factors matrix $F_{As}$ corresponding to the weights W2 in Table 7. The solid line is the mean over $P = 100$ runs. The dashed lines corresponding to one standard deviation from the mean in each direction are also included but are not visible in parts of this graph as the corresponding errors are relatively small; the errors are sizable in the beginning of the period shown.

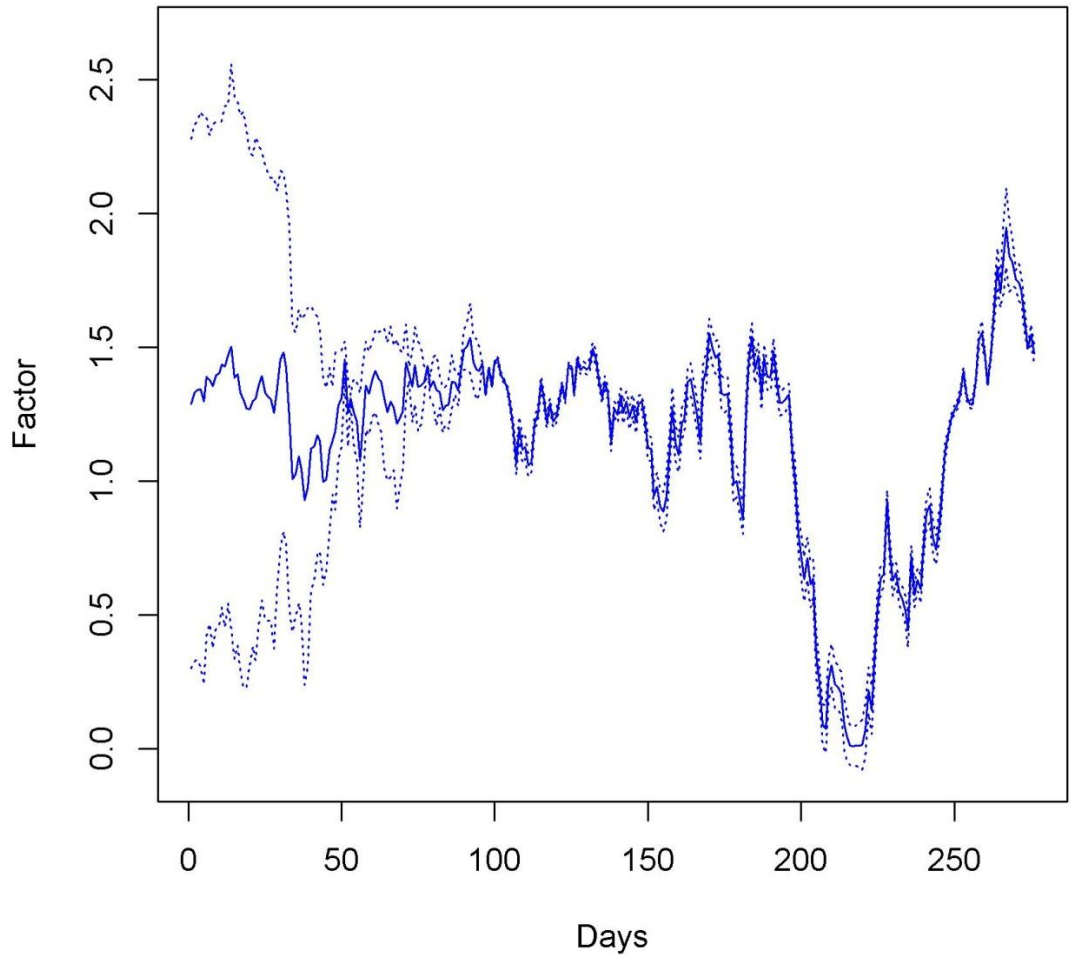

**Figure 17.** The time series plot for the third row of the factors matrix $F_{As}$ corresponding to the weights W3 in Table 7. The solid line is the mean over $P = 100$ runs. The dashed lines corresponding to one standard deviation from the mean in each direction are also included but are not visible in parts of this graph as the corresponding errors are relatively small; the errors are sizable in the beginning of the period shown.

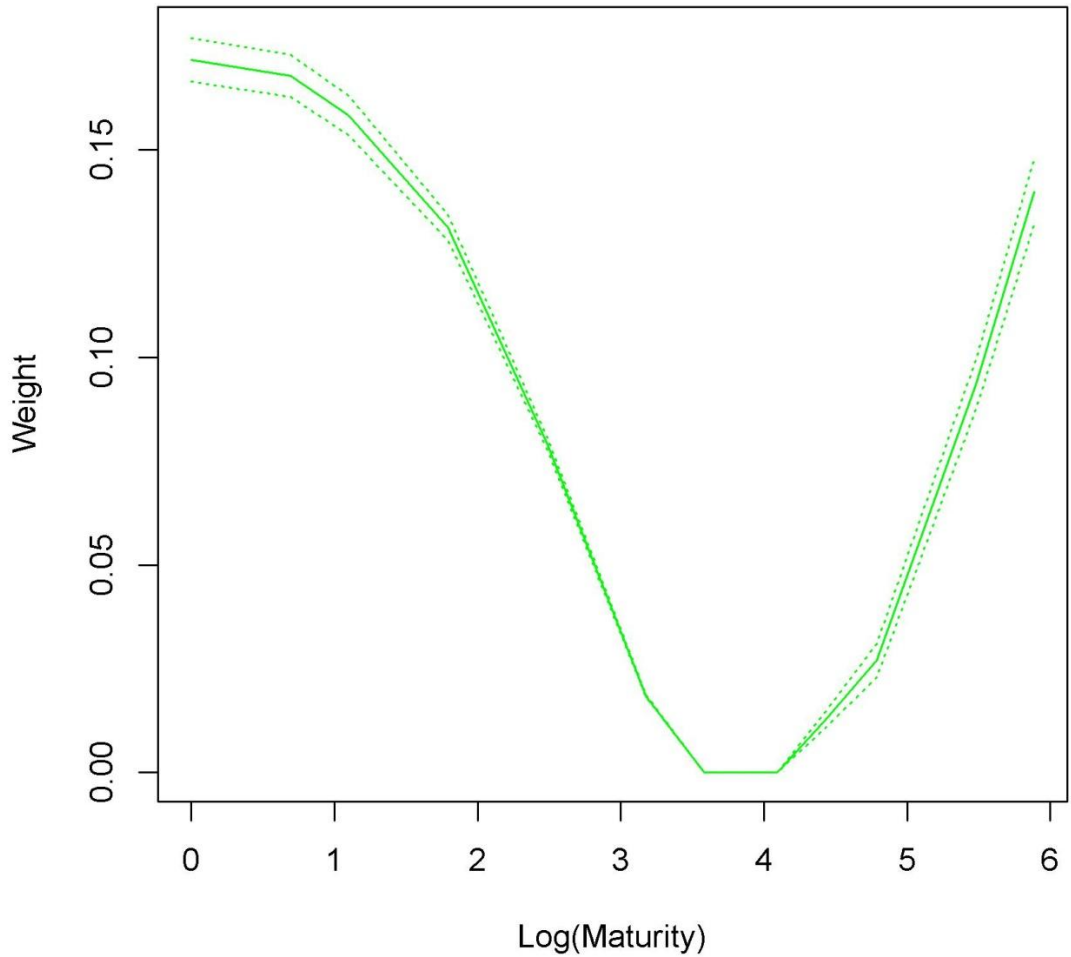

**Figure 18.** The weights W1 in Table 7 plotted against the natural log of the maturity. The solid line is the mean over $P = 100$ runs. The dashed lines correspond to one standard deviation from the mean in each direction.

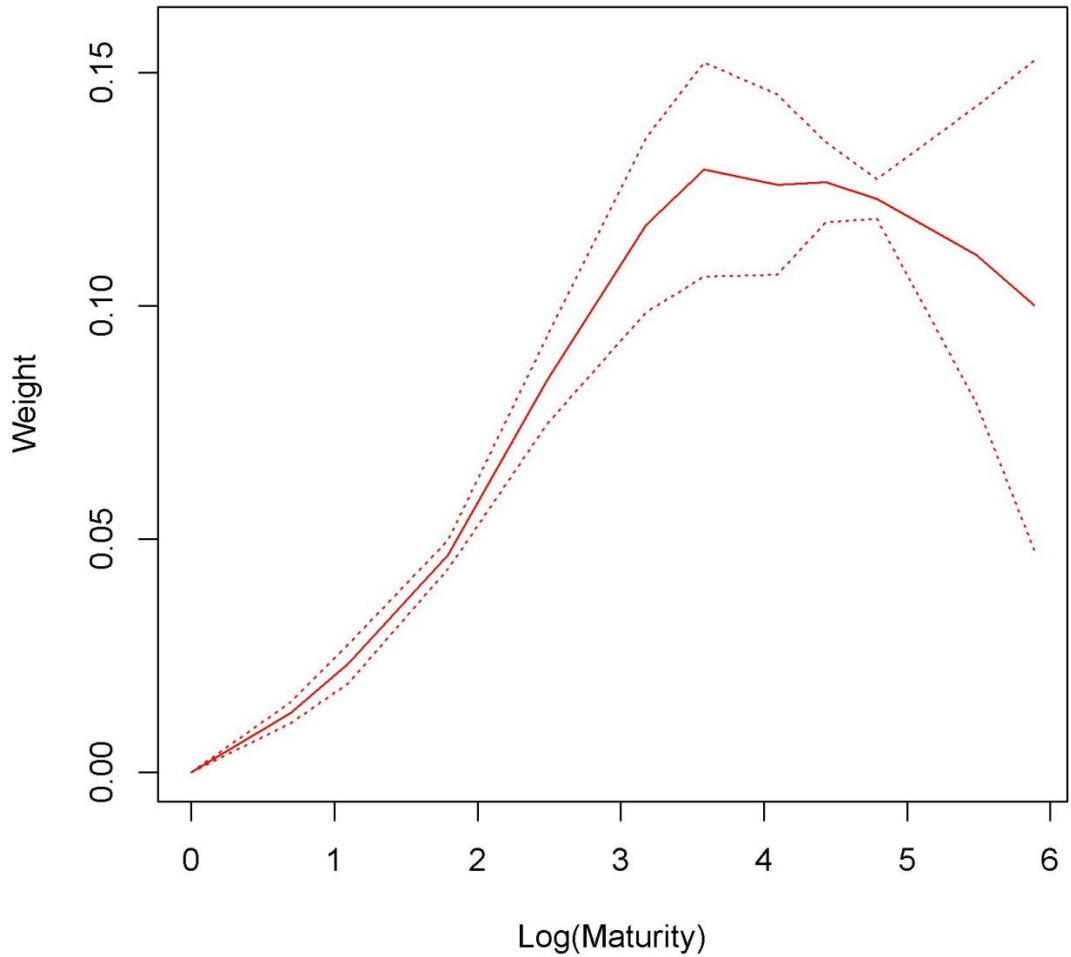

**Figure 19.** The weights W2 in Table 7 plotted against the natural log of the maturity. The solid line is the mean over $P = 100$ runs. The dashed lines correspond to one standard deviation from the mean in each direction.

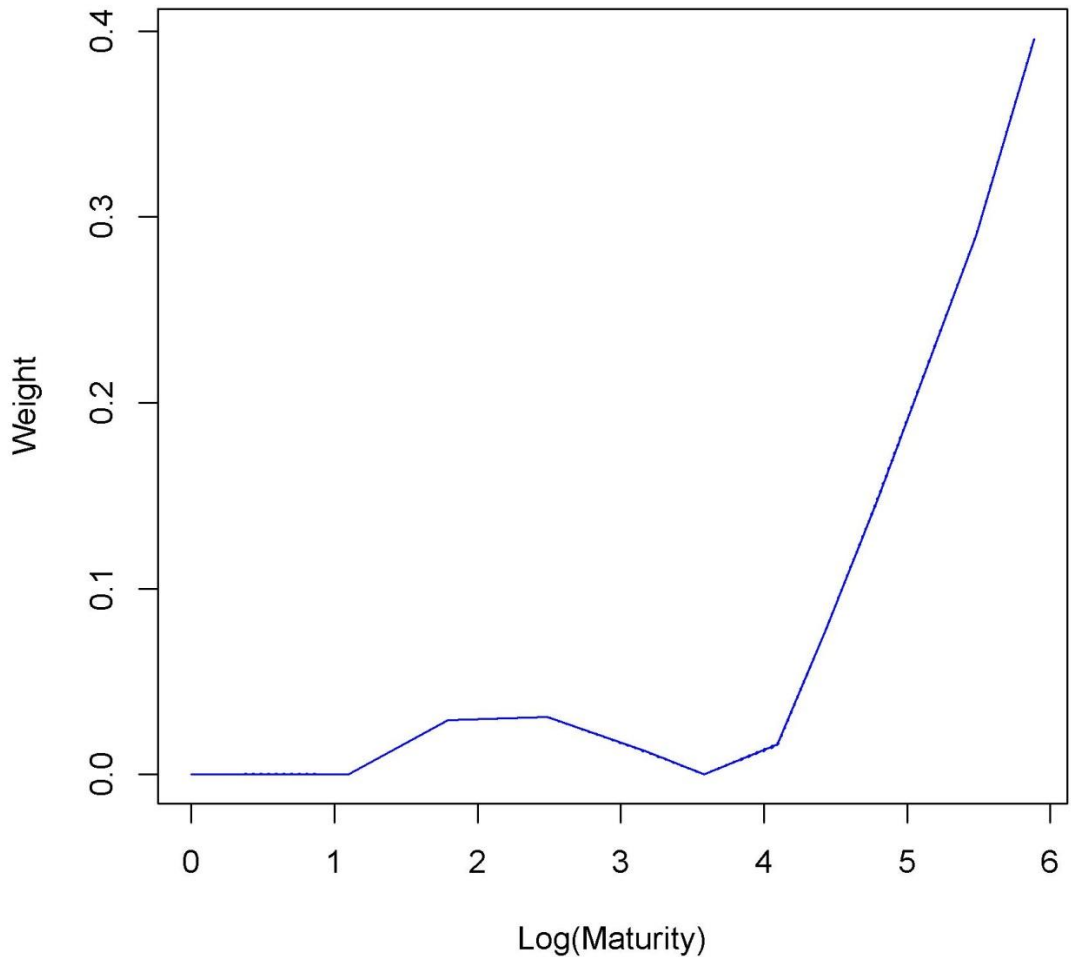

**Figure 20.** The weights W3 in Table 7 plotted against the natural log of the maturity. The solid line is the mean over $P = 100$ runs. The dashed lines corresponding to one standard deviation from the mean in each direction are also included but are not visible in this graph as the errors are relatively small (see Table 7).

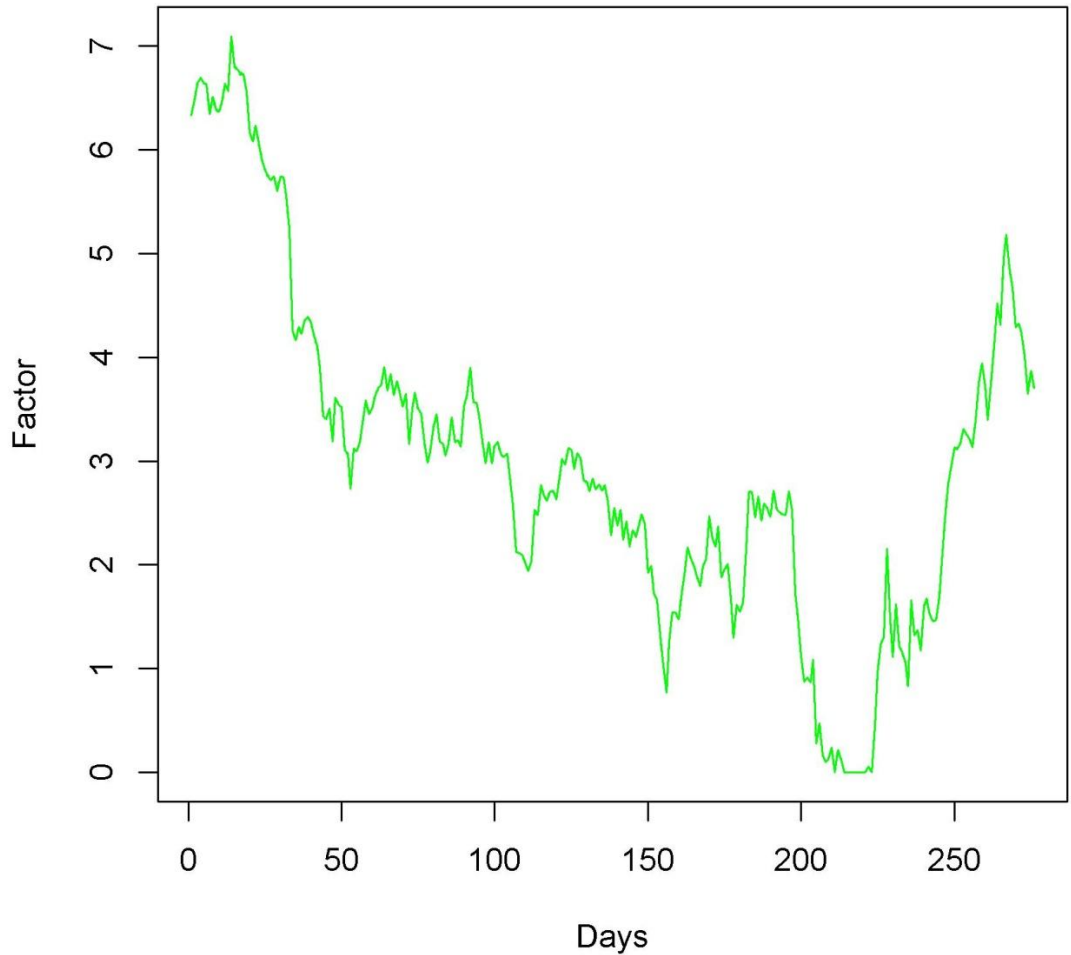

**Figure 21.** The time series plot for the first row of the factors matrix $F_{As}$ corresponding to the weights W1 in Table 10. The solid line is the mean over $P = 100$ runs. The dashed lines corresponding to one standard deviation from the mean in each direction are also included but are not visible in this graph as the errors are tiny (see Table 10).

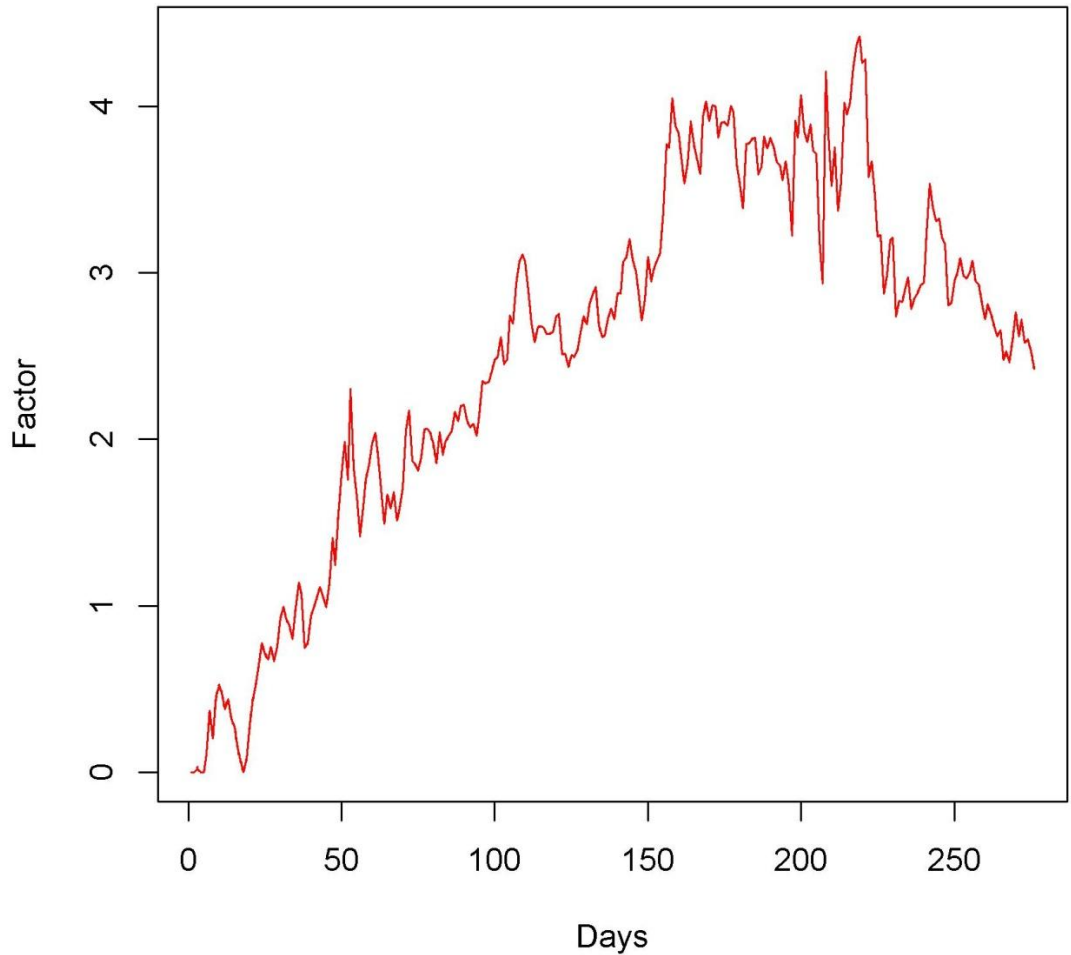

**Figure 22.** The time series plot for the second row of the factors matrix $F_{As}$ corresponding to the weights W2 in Table 10. The solid line is the mean over $P = 100$ runs. The dashed lines corresponding to one standard deviation from the mean in each direction are also included but are not visible in this graph as the errors are tiny (see Table 10).

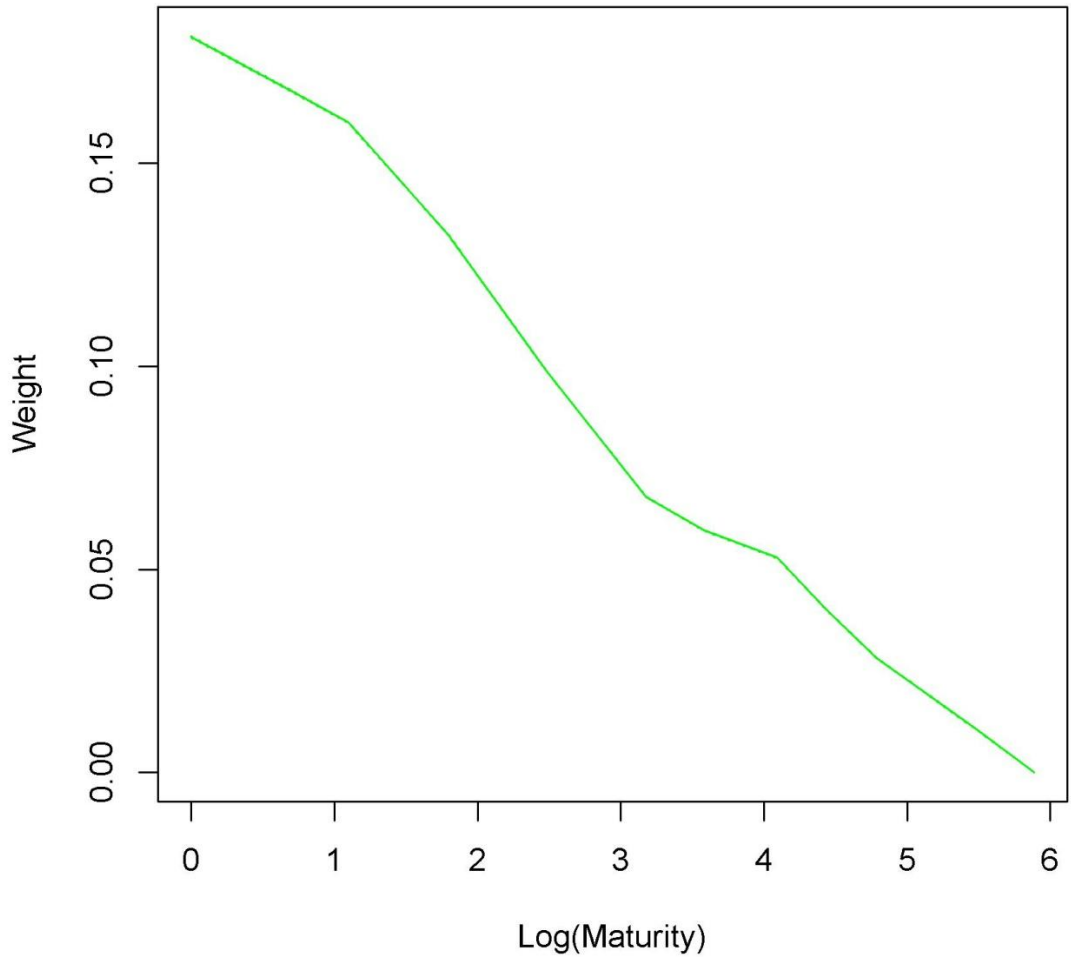

**Figure 23.** The weights W1 in Table 10 plotted against the natural log of the maturity. The solid line is the mean over $P = 100$ runs. The dashed lines corresponding to one standard deviation from the mean in each direction are also included but are not visible in this graph as the errors are tiny (see Table 10).

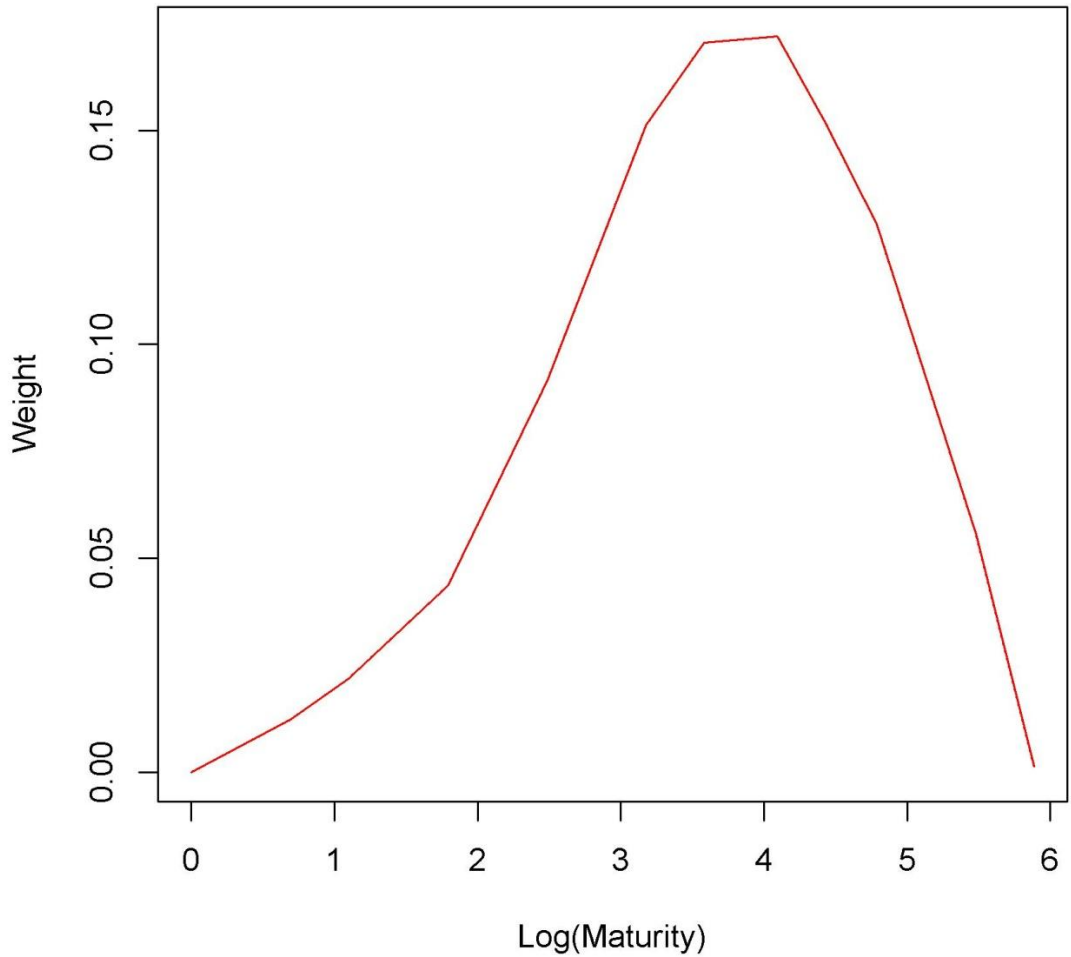

**Figure 24.** The weights W2 in Table 10 plotted against the natural log of the maturity. The solid line is the mean over $P = 100$ runs. The dashed lines corresponding to one standard deviation from the mean in each direction are also included but are not visible in this graph as the errors are tiny (see Table 10).

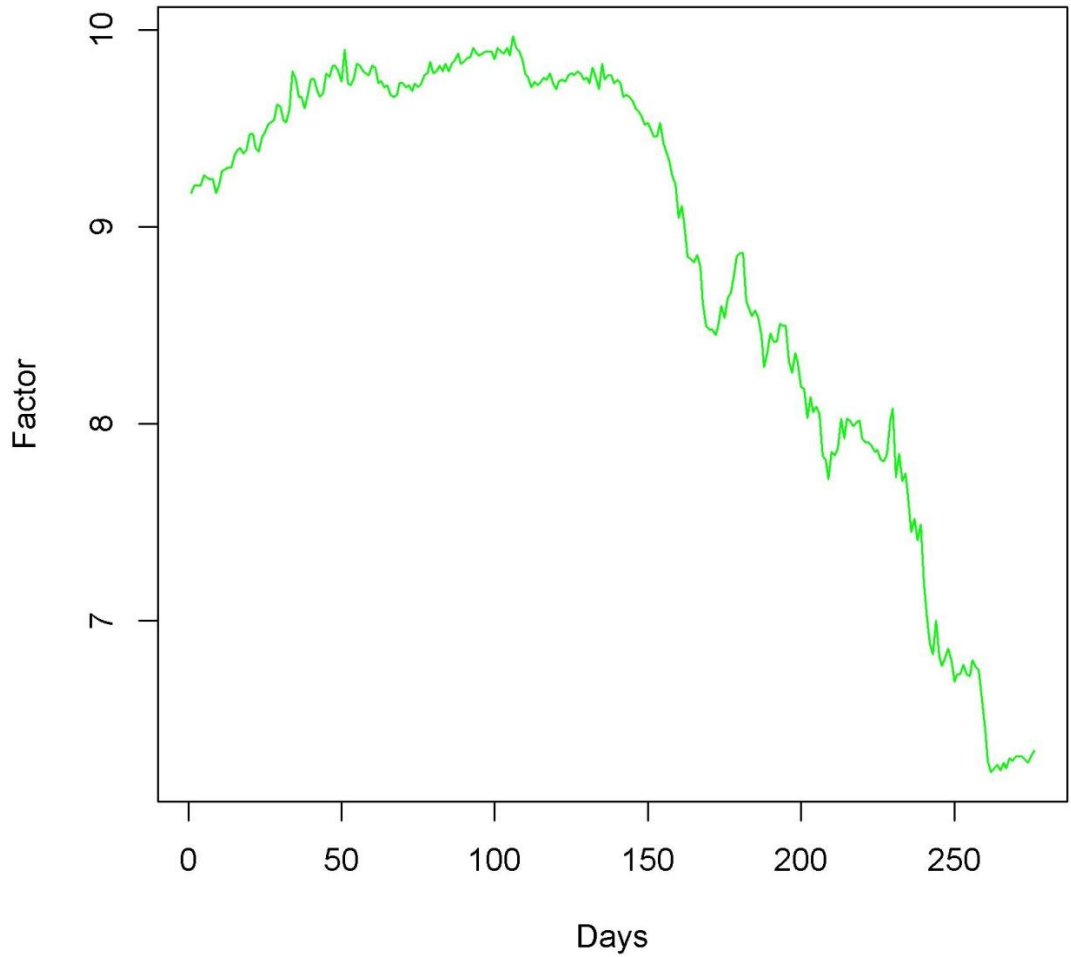

**Figure 25.** The time series plot for the first row of the factors matrix $F_{As}$ corresponding to the weights W1 in Table 13.

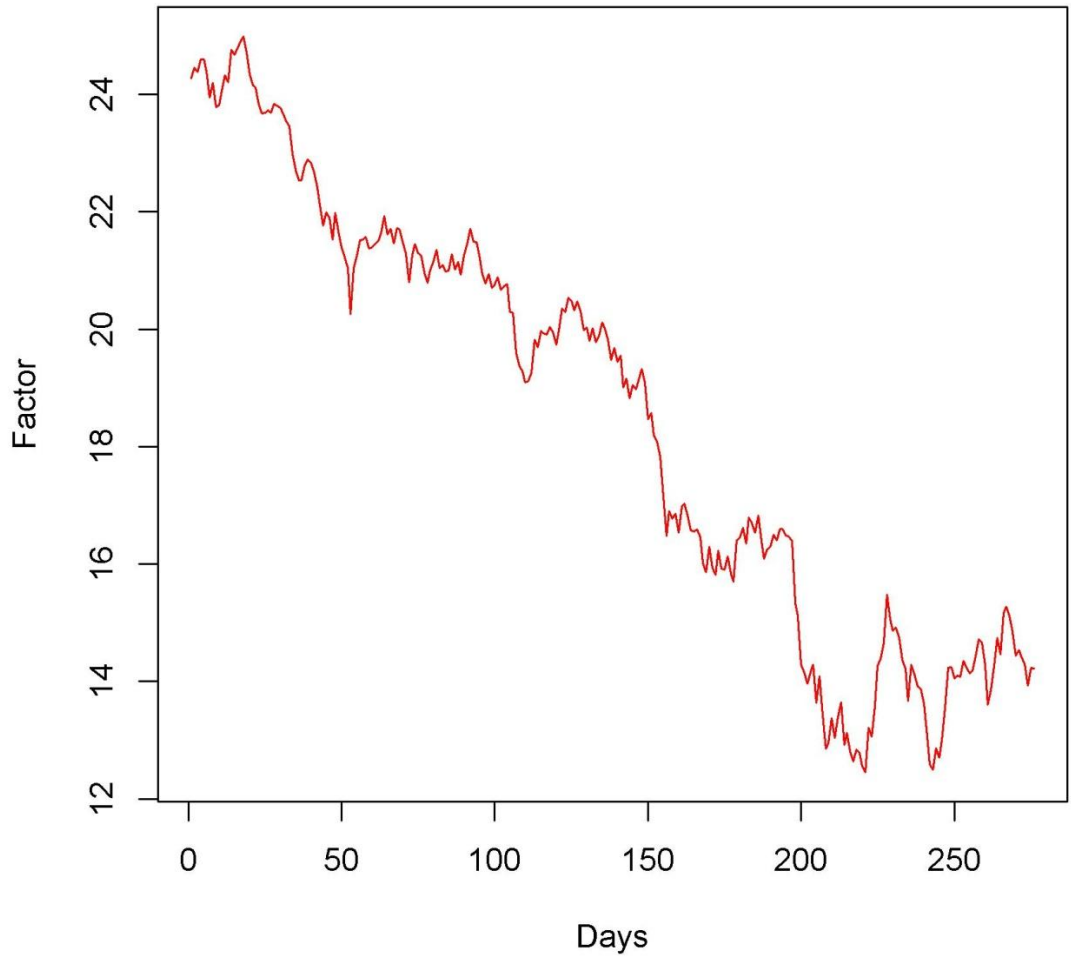

**Figure 26.** The time series plot for the second row of the factors matrix $F_{As}$ corresponding to the weights W2 in Table 13.

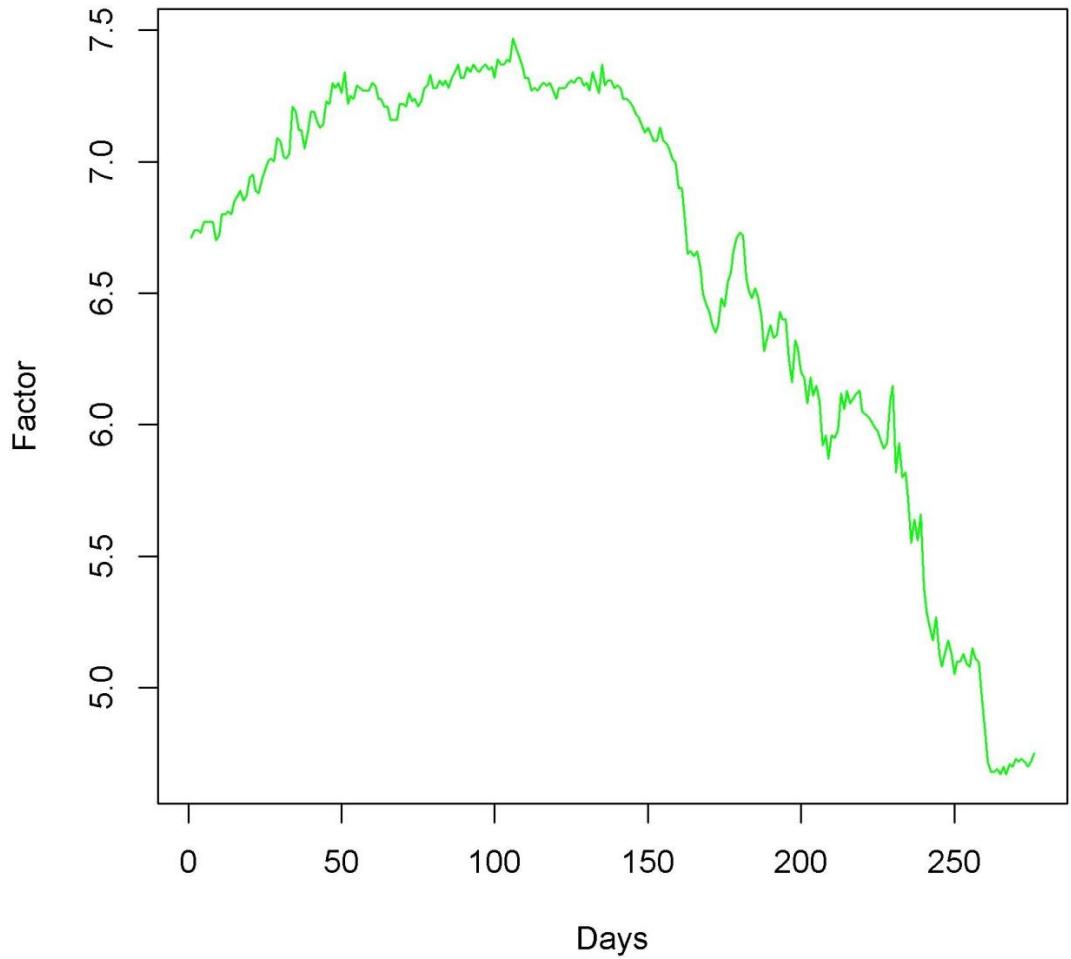

**Figure 27.** The time series plot for the first row of the factors matrix $F_{As}$ corresponding to the weights W1 in Table 15.

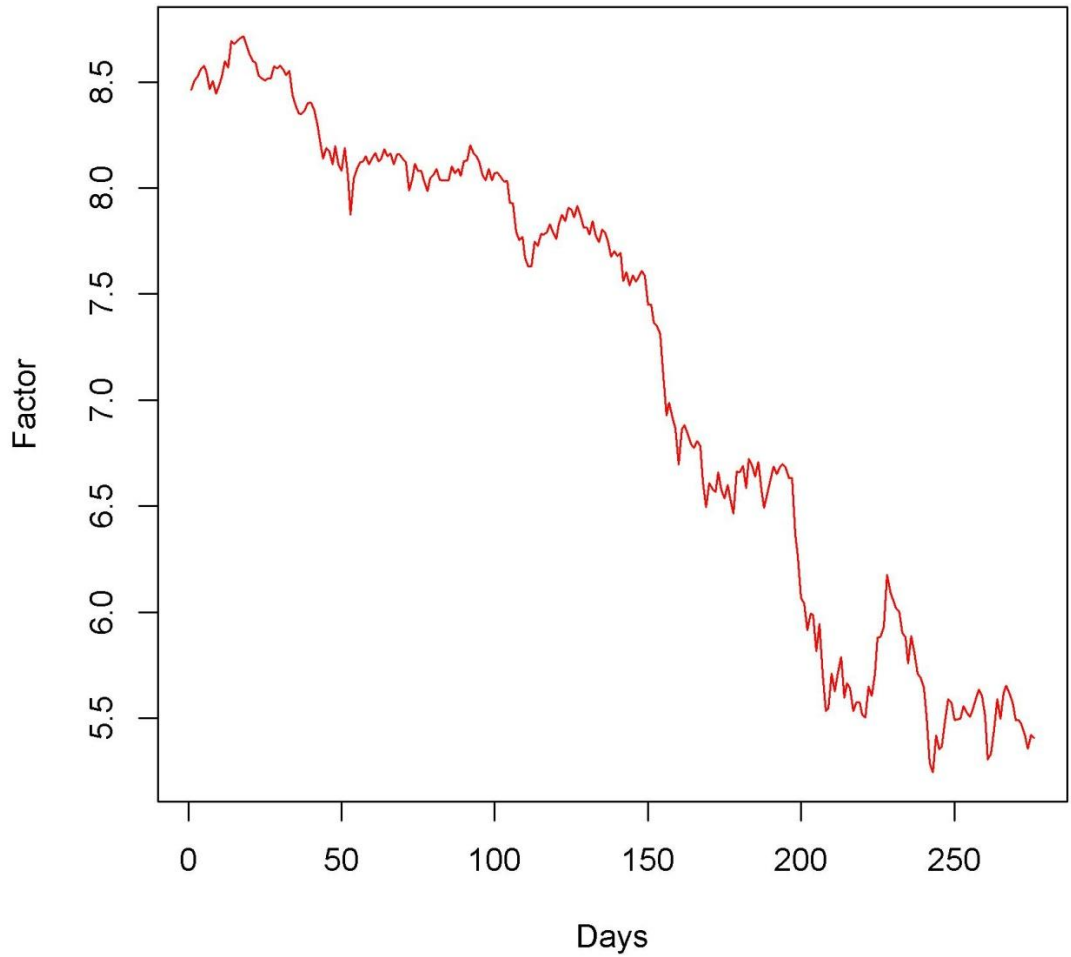

**Figure 28.** The time series plot for the second row of the factors matrix $F_{As}$ corresponding to the weights W2 in Table 15.

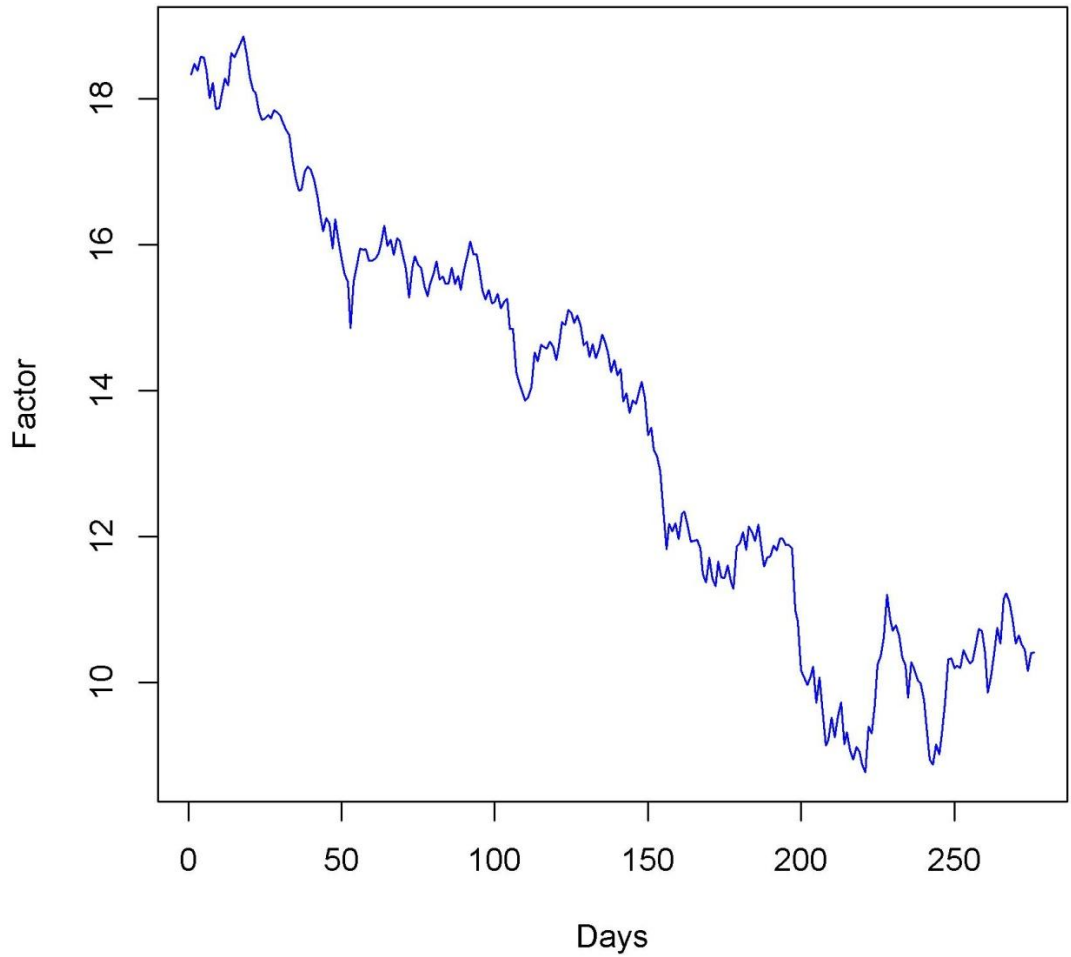

**Figure 29.** The time series plot for the third row of the factors matrix $F_{As}$ corresponding to the weights W3 in Table 15.

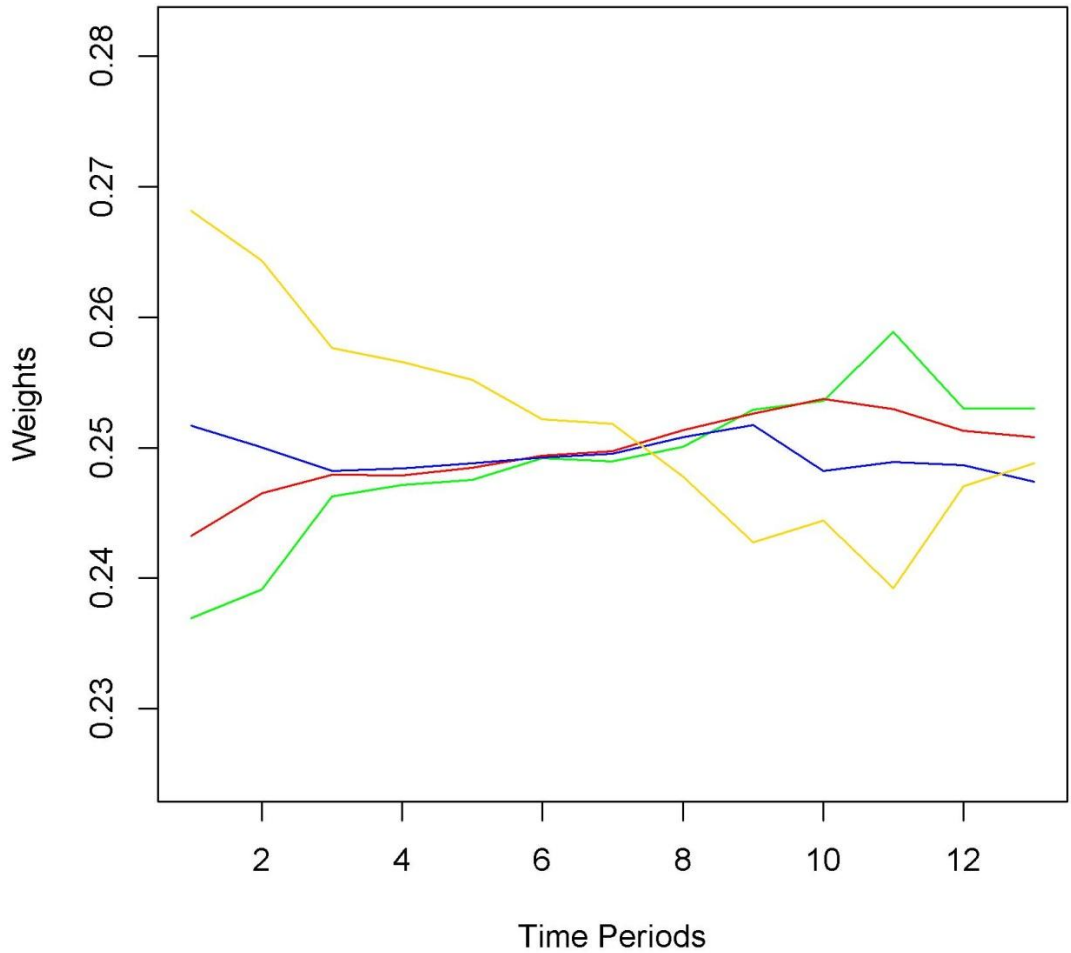

**Figure 30.** The first-cluster weights $W_{i1}$ ($i = 1, \ldots, 4$, which correspond to the maturities 1 Mo, 2 Mo, 3 Mo and 6 Mo) in the $K = 2$ cluster model (see Subsection 3.1) computed based on thirteen 21-trading-day periods (as opposed to the 276-trading-day period as in Table 13).

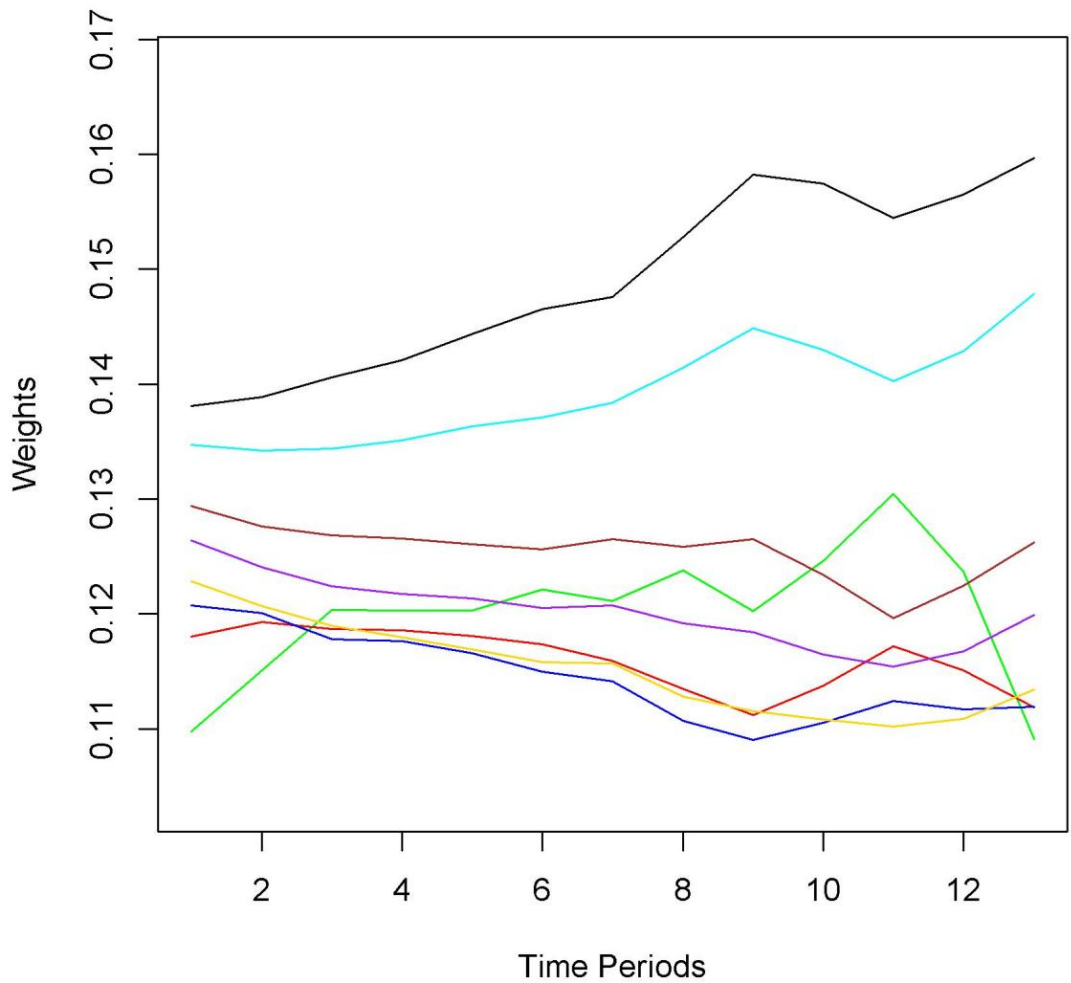

**Figure 31.** The second-cluster weights $W_{i2}$ ($i = 5, \ldots, 12$, which correspond to the maturities 1 Yr, 2 Yr, 3 Yr, 5 Yr, 7 Yr, 10 Yr, 20 Yr and 30 Yr) in the $K = 2$ cluster model (see Subsection 3.1) computed based on thirteen 21-trading-day periods (as opposed to the 276-trading-day period as in Table 13).

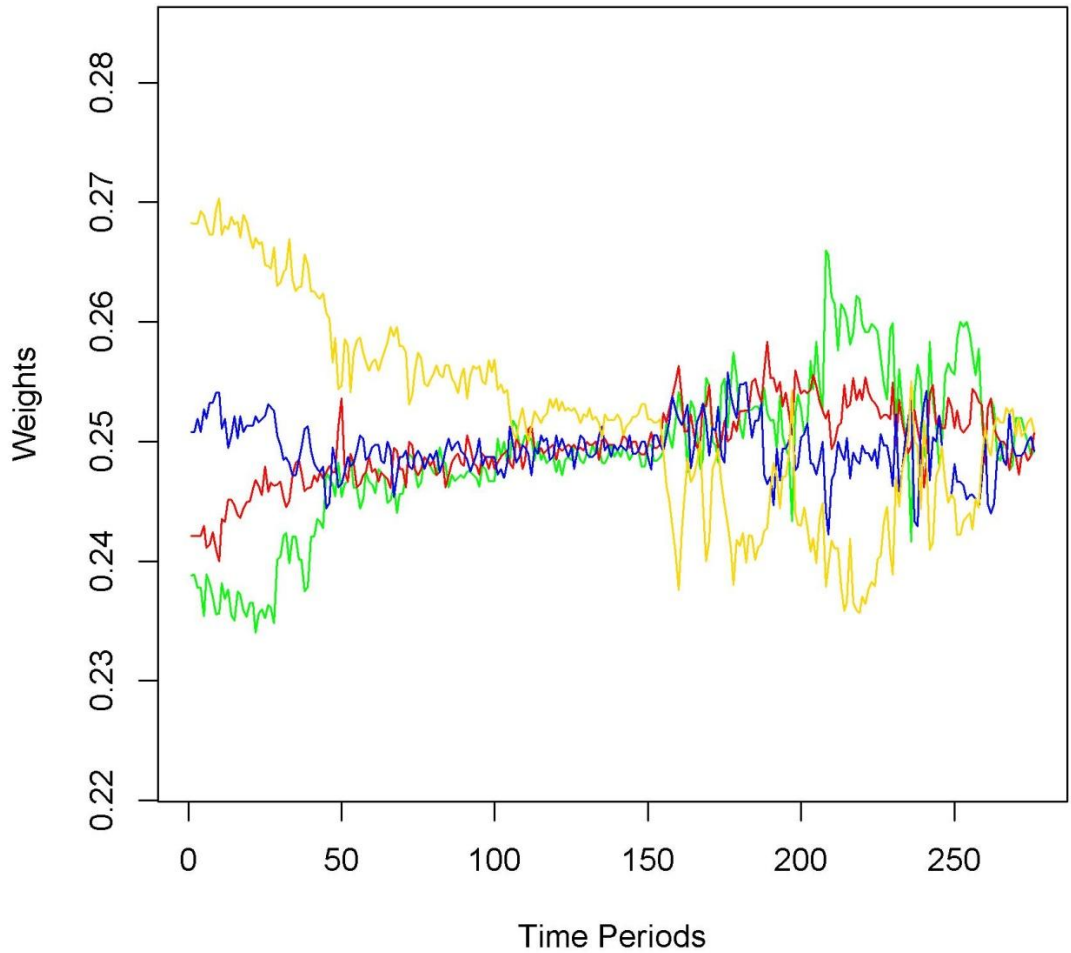

**Figure 32.** The first-cluster weights $W_{i1}$ ($i = 1, \ldots, 4$, which correspond to the maturities 1 Mo, 2 Mo, 3 Mo and 6 Mo) in the $K = 2$ cluster model (see Subsection 3.1) computed on each trading day (as opposed to the 276-trading-day period as in Table 13).

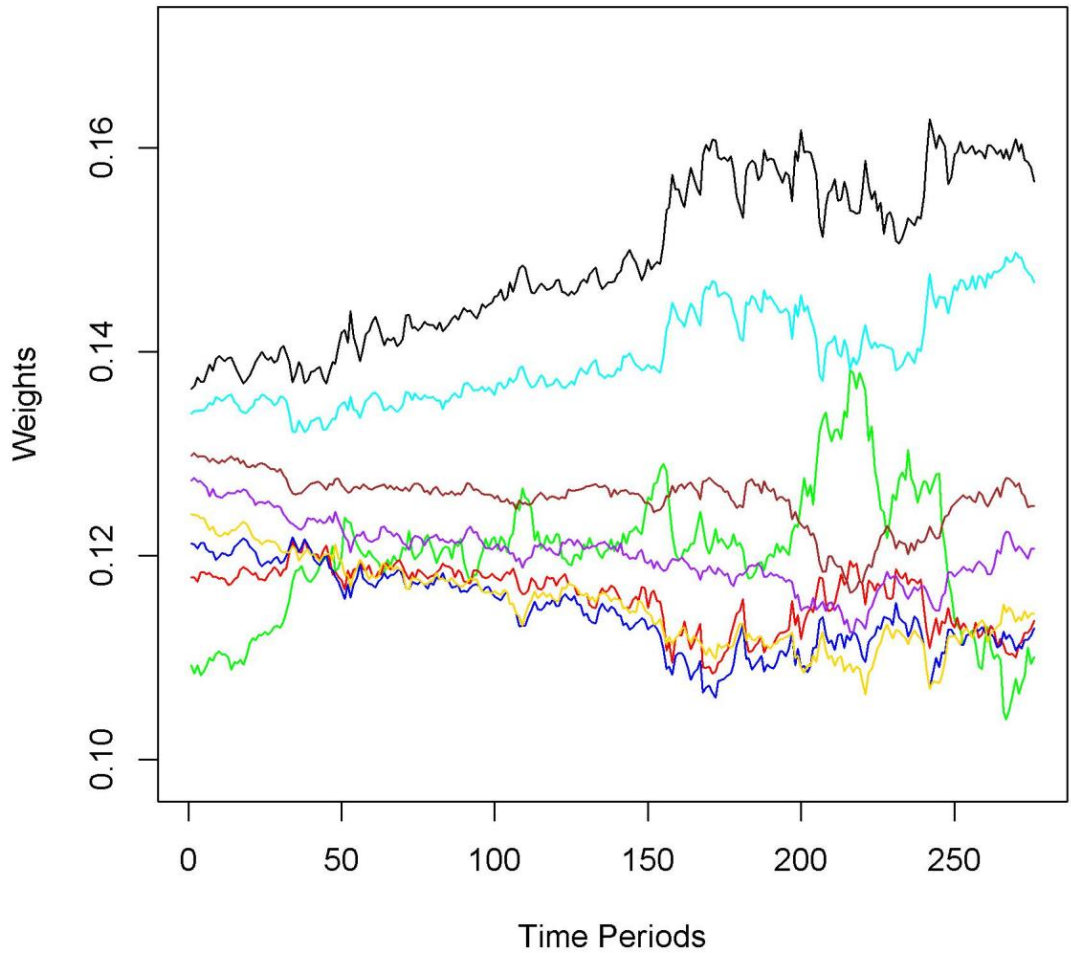

**Figure 33.** The second-cluster weights $W_{i2}$ ($i = 5, \ldots, 12$, which correspond to the maturities 1 Yr, 2 Yr, 3 Yr, 5 Yr, 7 Yr, 10 Yr, 20 Yr and 30 Yr) in the $K = 2$ cluster model (see Subsection 3.1) computed on each trading day (as opposed to the 276-trading-day period as in Table 13).